\documentclass[aps,prb,showpacs,twocolumn,floats,superscriptaddress,10pt]{revtex4-1}
\usepackage{graphicx}
\usepackage{dcolumn}
\usepackage{bm}
\usepackage{nicefrac}
\usepackage[latin3]{inputenc}
\usepackage[makeroom]{cancel}
\usepackage{amsmath}
\usepackage{amsfonts}
\usepackage{amssymb}
\usepackage{color}
\usepackage[left=2cm,right=2cm,top=2cm,bottom=2cm]{geometry}

\usepackage{simplewick}
\usepackage{array}
\usepackage{appendix}

\RequirePackage[
   hyperindex,colorlinks,bookmarksnumbered,
   plainpages=true,pdfstartview=FitH]{hyperref}
\hypersetup{linkcolor=blue,urlcolor=blue,citecolor=blue}
\usepackage{hyperref}

\definecolor{purple}{rgb}{0.5,0,0.6}

\definecolor{Jerome}{rgb}{0.7, 0.0, 0.3}

\usepackage{ulem}

\begin{document}

\title{Noises in a two-channel charge Kondo model}

\date{\today}

\author{T. K. T. Nguyen}
\email{nkthanh@iop.vast.vn}
\affiliation{Institute of Physics, Vietnam Academy of Science and Technology, 10 Dao Tan, 118000 Hanoi, Vietnam}
\author{J. Rech}
\email{jerome.rech@cpt.univ-mrs.fr}
\affiliation{Aix Marseille Universit\'e, Universit\'e de Toulon, CNRS, CPT, Marseille, France}
\author{T. Martin}
\affiliation{Aix Marseille Universit\'e, Universit\'e de Toulon, CNRS, CPT, Marseille, France}
\author{M. N. Kiselev}
\affiliation{The Abdus Salam International Centre for Theoretical Physics, Strada
Costiera 11, I-34151, Trieste, Italy}


\begin{abstract}
We investigate fluctuations of electric and heat currents, along with their cross-correlations, in a two-channel charge Kondo circuit driven by either a voltage bias or a temperature gradient applied across the weak link. The ratios of voltage-driven electric/heat noise to the applied voltage $V$ exhibit oscillations with the gate voltage $N$, resembling the behavior of the thermoelectric coefficient $G_T$. In contrast, the ratios of temperature-driven electric/heat noise to the temperature difference $\Delta T$ vary with $N$ in a manner analogous to the thermal coefficient $G_H$ or the electric conductance $G$. The mixed noise, which is defined as the correlation function between electric and heat currents, displays behavior opposite to that of the above noises. The logarithmic temperature dependence of these noises signals non-Fermi-liquid behavior, while their oscillations with gate voltage reflect the roles of particle-hole and time-reversal symmetries in thermoelectric transport. Our results demonstrate that the fundamental relations linking voltage- and temperature-induced noises to thermoelectric transport across a tunnel junction persist beyond the Fermi-liquid paradigm.
\end{abstract}

\maketitle

\section{Introduction} 

Thermoelectricity has become a significant topic in modern physics, driven by the growing demand for advanced thermoelectric materials~\cite{Wood,TEbook1,TEbook2}. One of the primary approaches in this search is the development of nanostructured devices~\cite{TE_nano}. Recent advances in the fabrication of these devices have opened up new possibilities for exploring a wide range of quantum phenomena related to charge, spin, and thermoelectric effects~\cite{Blanterbook, Kiselevbook}. The impact of quantum effects on the properties of quantum devices is generally studied initially in the single-electron transistor (SET) due to its ability to be precisely controlled by external fields, such as electric potential and magnetic fields~\cite{Blanterbook, Kiselevbook}. SET devices provide valuable insights into the effects of strong electron interactions, interference, and resonant scattering on quantum transport.

A fundamental phenomenon that encapsulates both resonant scattering and strong interactions is the Kondo effect~\cite{Kondo, Hewson} where a local spin is coupled to conduction electrons. The Kondo physics in SET is realized by transport measurements~\cite{Kondoreview}.  In the ground state of the single impurity $S =1/2$ single channel Kondo model (1CK), the spin of the dot is screened by a cloud of conduction electrons, forming a singlet state. The low-energy behavior of the remaining electrons is characterized by a local Fermi liquid (FL) theory~\cite{Nozieres}. Electrons are scattered by the singlet both elastically and inelastically. The ratio of elastic to inelastic scattering is fixed by universality. The Kondo temperature $T_K$ is the only scale that governs the low-energy properties of the model. However, the behavior of the $M$-orbital spin-$S$ Kondo model at energies below the Kondo temperature $T_K$ depends on how the mobile electrons screen the impurity spin. The system exhibits coherent behavior and FL properties when the system is fully or underscreened ($M\le 2S$) while it is likely to exhibit non-Fermi liquid (NFL) characteristics in the overscreened case ($M >2S$)~\cite{NozieresBlandin}. However, it is difficult to achieve the strong NFL regime at very low temperature, namely $T\ll T^* < T_K$, where $T^*$ is related to the perturbative expansion parameter $|r|$ as $T^*=|r|^2 T_K$. The system has a tendency to fall into the FL regime associated with the stable FL fixed point~\cite{thanh2010}. Nevertheless, in higher temperature regimes, $T^* < T < T_K$, the fingerprints of the weak NFL behavior can be observed~\cite{thanh2010}.

Going beyond conventional knowledge that the Kondo effect is attributed to the spin degrees of freedom of the quantum impurity, many unconventional Kondo phenomena have been observed in a variety of systems~\cite{unKondo1,unKondo2,unKondo3,unKondo4}. For instance, the charge Kondo effect deals with an iso-spin implementation of the charge quantization~\cite{flensberg,matveev,furusakimatveev}. This charge Kondo model has been implemented in pioneering breakthrough experiments that involve the edge currents of the integer quantum Hall effect~\cite{pierre2,pierre3}. These experiments mark a significant step forward in the study of multi-channel charge Kondo effects~\cite{michellprl,thanhprl}. Indeed, fairly recently, another experimental study~\cite{2SCKC} has successfully implemented a tunable nanoelectronic circuit consisting of two coupled hybrid metallic-semiconductor islands.

Recently, thermoelectric transport through quantum dot (QD) systems has garnered significant attention from both theorists~\cite{Beenakker,Turek,MAprl,MAprb,Krawiec,Costi,Trocha,Donsa,Wojcik} and experimentalists~\cite{Staring,TE_exp1,TE_exp2,TE_exp3,Svilans,Dutta}. Measurements of thermoelectric coefficients are challenging because they require heating the contacts~\cite{thanhprl,MAprl,MAprb,thanh2010,karki2020}. Keeping the temperature drop small and controllable during measurements is particularly difficult for experimentalists. To address this, it is essential to relate the thermoelectric coefficients using established relations, such as the Cutler-Mott formula~\cite{CM}, which connects thermopower to electrical conductance~\cite{thanharx}, and the Wiedemann-Franz law~\cite{WF}, which links thermal conductance to electrical conductance~\cite{kiselev2023}.

Furthermore, since current noise measurement is known to provide valuable insights into the fundamental mechanisms of quantum transport and electron interactions\cite{shotnoise1,shotnoise2,shotnoise3}, it is also a promising experimental method for investigating the thermoelectric properties of Kondo problems~\cite{pierre2}. Moreover, noise measurements are capable of probing the out-of-equilibrium properties of the model~\cite{LandauSela,reviewGiazotto}, a field that has not been extensively studied due to the limited number of theoretical approaches~\cite{Meir2002,Golub,Gogolin} available for out-of-equilibrium situations. 

Among the various types of noise observed in mesoscopic systems, shot noise -- arising from the discrete nature of charge carriers -- plays a crucial role, particularly in systems where transport is dominated by tunneling events [see the review~\cite{shotnoisereview} and references therein]. Recently, delta-T noise, associated with temperature differences, has garnered significant attention from physicists~\cite{Tnoise_exp1,Tnoise_exp2,Tnoise_exp3,Tnoise_exp4,Tnoise_exp5,Tnoise_exp6,Tnoise_theo1,Tnoise_theo2,Tnoise_theo3,Tnoise_theo4,Tnoise_theo5}. This type of noise is expected to serve as a new probe of quantum effects that cannot be observed through shot noise measurements. Delta-T noise, in particular, provides unique insights into the interplay between thermal effects and quantum correlations in transport phenomena. Notably, both shot noise and delta-T noise at low temperatures have the potential to yield valuable information about the statistical nature of charge transfer~\cite{detect_exp1,detect_exp2,Tnoise_theo5}. Furthermore, FL interactions can be extracted from the measurement of shot noise in a $SU(N)$ Kondo quantum dot\cite{LeHur}. In this context, we investigate whether noise can probe the NFL characteristics in a multi-channel charge Kondo model.

\begin{figure}[t]
\begin{center}
\includegraphics[scale=0.24]{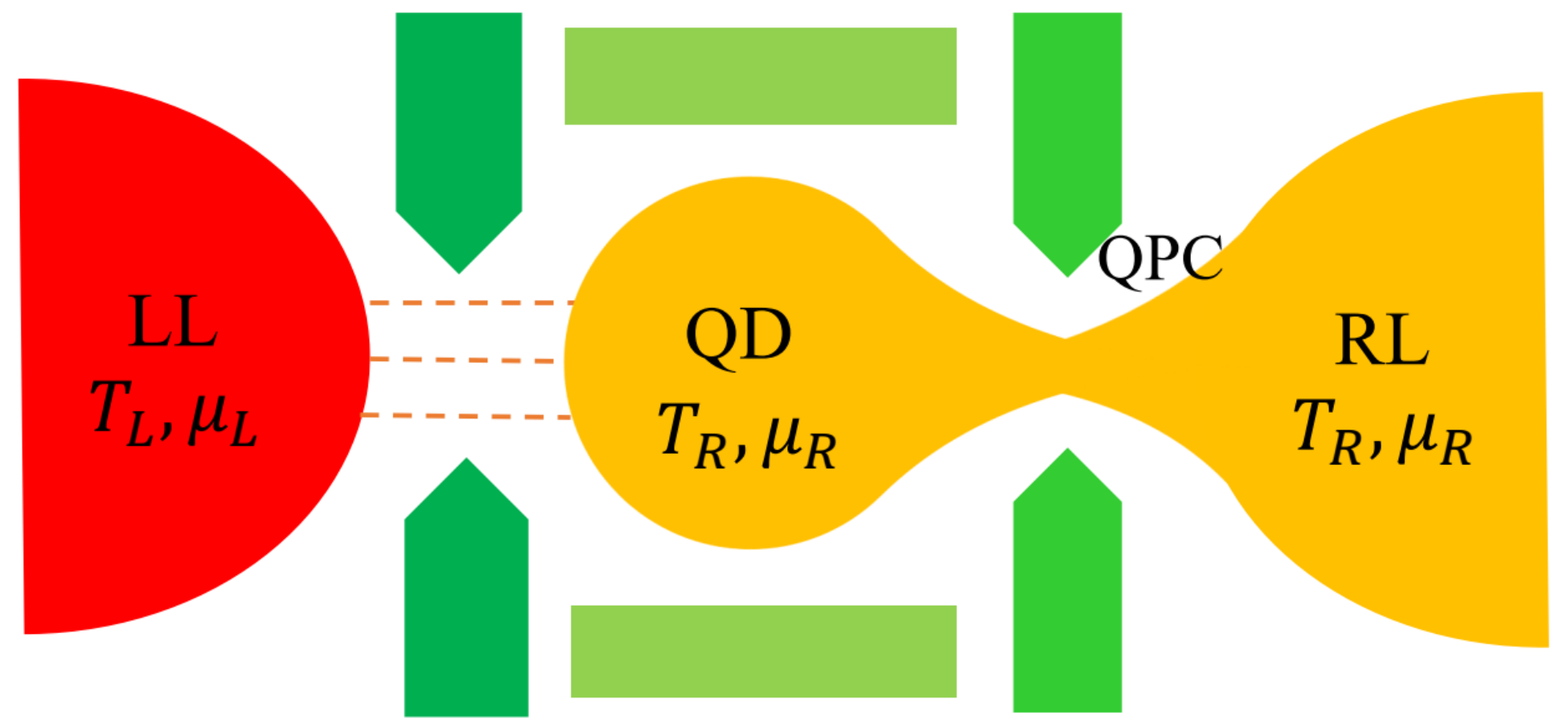}
\caption{Schematic of a single-electron transistor device in which a large metallic quantum dot (QD) is embedded into two-dimensional electron gas (2DEG) and connects weakly to the left lead (LL) through a tunnel barrier and strongly coupled to the right one (RL) through an almost transparent single-mode quantum point contact (QPC). The QD and the right lead (yellow color) are at potential $\mu_R$ temperature $T_R=T$ while the left lead (red color) is at higher voltage $\mu_L=\mu_R-eV$ and temperature $T_L=T +\Delta T$. The voltage or temperature drops at the weak link. The green patches demonstrate the gate voltage and the voltages controlling the tunnel barrier and the QPC.}
\label{fig1}
\end{center}
\end{figure}

In this work, we calculate the voltage-driven electric/heat noise, temperature-driven electric/heat noise, and voltage/temperature-driven mixed noise (charge-heat cross correlations) of the electric and heat currents at the weak link connecting a reservoir to a two-channel charge Kondo (2CK) circuit \cite{comment}. The setup, illustrated in Fig.~\ref{fig1}, consists of a large metallic quantum dot (QD) that is weakly coupled to the left lead via a tunnel barrier and strongly coupled to the right lead through an almost transparent single-mode quantum point contact (QPC)~\cite{flensberg,matveev,furusakimatveev}. The key idea behind mapping this system to a 2CK problem lies in treating the two degenerate charge states of the QD as a ``quantum impurity'', while the electrons' position inside or outside the dot defines an isospin degree of freedom. The electrons' spin projections then naturally define the two screening channels in the Kondo model~\cite{MAprl,MAprb}.

As discussed in Refs.~\cite{MAprl,MAprb,thanh2010,thanhcom2023,thanh2024}, perturbative approaches, based on the assumption of weak backscattering at the QPCs ($|r|\ll 1$), are valid in the temperature regime $|r|^2 T_K\ll T\ll T_K$. Moreover, our model also allows for a nonperturbative treatment of $|r|$. Specifically, we compute the relevant correlation functions without assuming small reflection amplitudes, even though the charge Kondo model itself requires small $|r|$. As a result, our results extend beyond the perturbative regime and remain valid even at temperatures $T\le |r|^2 T_K$.

Electric and heat currents, as well as their associated noises (electric, heat and mixed noise), are computed up to the linear order in voltage or temperature bias -- that is, we retain only the first-order terms of the expansions in $eV$ or $\Delta T$. We find that the voltage-driven electric and heat noises show oscillatory behavior with respect to the gate voltage $N$, mirroring the pattern observed in the thermoelectric coefficient $G_T$. On the other hand, the temperature-driven electric and heat noises display a dependence on $N$ similarly to that of the thermal coefficient $G_H$ or the electric conductance $G$. In contrast, the mixed noise in both situations exhibits a trend opposite to that of the aforementioned noises. 

Notably, the logarithmic temperature dependence of voltage-driven electric and heat noises and temperature-driven mixed noise in the vicinities of Coulomb peaks provides clear evidence of the NFL behavior characteristic of the 2CK state. The distinctive behavior of the Fano factors further reinforces the connection between current noises and thermoelectric coefficients. {\color{black} Comparisons with the single-channel Kondo (1CK) model across all noise types confirm the crossover flow from the 2CK to the 1CK regime.}

The paper is organized as follows. We describe the theoretical model in Sec. II. Equations for the currents and noises are presented in Sec. III while the corelation function and density of states are introduced in Sec. IV. The main results are demonstrated and discussed in Sec. V. We conclude our work in Sec. VI.

\section{Model} 
We analyze a SET device, as illustrated in Fig.~\ref{fig1}~\cite{MAprl,MAprb,thanh2010,thanhcom2023}. The central component consists of a large metallic QD operating in the weak Coulomb blockade (CB) regime. This QD is strongly coupled to the right lead through a nearly transparent single-mode QPC. The QD-QPC structure is embedded within a two-dimensional electron gas (2DEG). We assume that the QD-QPC structure is in thermal equilibrium at a potential $\mu_R$ and a temperature $T_R=T$ (represented by the yellow region, the right electrode, in Fig.~\ref{fig1}). The left lead, also part of the 2DEG, is connected to the QD via a tunnel barrier. It is assumed to be at a higher temperature $T_L=T+\Delta T$ (shown as the red region in Fig.~\ref{fig1}, with a potential of $\mu-eV$). This setup is used to investigate the voltage and temperature bias-driven noises of electric and heat currents at the tunnel barrier. The temperature difference across the tunnel barrier, $\Delta T$, is controlled using a current heating technique~\cite{TE_exp2}. We assume that $\Delta T$ is small relative to the reference temperature $T$, which allows us to apply a perturbative expansion in terms of $V$ and $\Delta T$.

The Hamiltonian describing the system in which the QD-QPC structure is weakly coupled to the left lead, is given by $H = H_L + H_R + H_T$. Here, $H_L= \sum_{\mathbf{k},\alpha}\epsilon_{\mathbf{k},\alpha}c^\dagger_{\mathbf{k},\alpha}c_{\mathbf{k},\alpha}$ describes the nonineracting left lead, ($c$ denotes the electrons in the left lead, $\alpha=\uparrow,\downarrow$ stands for the two spin projections of electrons), while $H_R$ describes the right part of the junction, namely the interacting quantum dot strongly coupled to the right lead. The tunneling from the left lead to the dot is given by the tunneling Hamiltonian $H_T$ which reads
\begin{equation}
H_T = \sum_{\mathbf{k},\alpha} t_{\mathbf{k},\alpha} c^\dagger_{\mathbf{k},\alpha} d_\alpha + \text{H.c.} ,
\label{eq:tunnelH}
\end{equation}
where $d$ corresponds to the electrons in the dot, and $t_{\mathbf{k},\alpha}/D\ll 1$ represents the tunneling amplitude ($D$ being the bandwidth).

The QD-QPC structure is described by the following Hamiltonian 
\begin{eqnarray}
H_R&=&\sum_\alpha\epsilon_\alpha d^\dagger_\alpha d_\alpha\nonumber\\
&& +\sum_\alpha\frac{v_{F}}{2\pi}\!\!\int^{\infty}_{-\infty}\!\!\!\!\left\{\left[\Pi_\alpha\left(x\right)\right]^2+\left[\partial_x\phi_\alpha\left(x\right)\right]^2\right\}dx\nonumber\\
&&+E_C\left[\hat{n}+\frac{1}{\pi}\sum_\alpha\phi_\alpha\left(0\right)-N\left(V_g\right)\right]^2\nonumber\\
&& -\frac{D}{\pi}\sum_\alpha|r_\alpha|\cos\left[2\phi_\alpha\left(0\right)\right],
\label{HR}
\end{eqnarray}
where $v_{F}$ is the Fermi velocity, $\phi_\alpha$ is a bosonic displacement operator describing transport through the QPC with a scatterer at $x=0$, and $\Pi_\alpha$ is the conjugated momentum $\left[\phi_\alpha\left(x\right),\Pi_{\alpha^\prime}\left(x^\prime\right)\right]=i\pi\delta\left(x-x^\prime\right)\delta_{\alpha,\alpha^\prime}$~\cite{Giamarchi}. The operator $d_\alpha$ is related to the fermionic field as $d_\alpha=\psi_\alpha\left(-\infty\right)$ with $\psi_\alpha\left(x\right)\sim e^{i\phi_\alpha\left(x\right)}$ in the one-dimensional model that describes the QPC. The third term describes the Coulomb interaction in the dot where $E_C=e^2/2C$ is the charging energy ($C$ is the QD capacitance), $N\left(V_g\right)=CV_g/e$ is a dimensionless parameter which is proportional to the gate voltage $V_g$, $\hat{n}$ is the operator depicting the number of electrons entering the QD through the left tunnel barrier, while $\sum_\alpha\phi_\alpha\left(0\right)/\pi$ is the number of electrons entering through the QPC (from the right lead)~\cite{Aleiner}. The last term demonstrates the backward scattering in the QPC with the small reflection amplitudes $|r_\uparrow|=|r_\downarrow|=|r|$ for the symmetric 2CK (no magnetic field is applied perpendicularly to the 2DEG plane).

\section{General formulas for currents and noises} 

The device under consideration is out-of-equilibrium at the tunneling barrier [see Fig.~\ref{fig1}]. The resulting operator for the current flowing through this barrier is given by $I_C = e d N_L/d t$$ = i e \left[ H_T , \sum_{\mathbf{k},\alpha} c^\dagger_{\mathbf{k},\alpha} c_{\mathbf{k},\alpha} \right]$$ = -i e \sum_{\mathbf{k},\alpha} \left( t_{\mathbf{k},\alpha} c^\dagger_{\mathbf{k},\alpha} d_\alpha - t^*_{\mathbf{k},\alpha} d^\dagger_\alpha c_{\mathbf{k},\alpha}  \right)$, recovering the standard expression for the current through a point-like junction between two reservoirs. Similarly, the heat current flowing through the tunnel barrier can be defined as $I_Q=d\left(H_L-\mu_L N_L\right)/dt$$= -i  \sum_{\mathbf{k},\alpha} \left( \epsilon_\mathbf{k} - \mu_L \right) \left( t_{\mathbf{k},\alpha} c^\dagger_{\mathbf{k},\alpha} d_\alpha - t^*_{\mathbf{k},\alpha} d^\dagger_\alpha c_{\mathbf{k},\alpha}  \right)$.

We consider the averages of both the charge current $I_C$ and the heat current $I_Q$ across the tunnel contact where a voltage bias $V$ and temperature drop $\Delta T$ are applied~\cite{Onsager}:
\begin{equation}
\left(\begin{array}{c}
\langle I_C \rangle\\
\langle I_Q \rangle
\end{array}\right)=\left(\begin{array}{cc}
G & G_{T}\\
TG_{T} & G_{H}
\end{array}\right)\left(\begin{array}{c}
 V\\
\Delta T
\end{array}\right),
\label{OnsagerRelation}
\end{equation}
where $G=\partial \langle I_C \rangle/\partial V|_{\Delta T=0}$, $G_T=\partial \langle I_C \rangle/\partial \Delta T|_{V=0}$, $G_H=\partial \langle I_Q \rangle/\partial \Delta T|_{V=0}$ are the electric conductance, thermoelectric coefficient, and thermal coefficient, respectively.

\subsection{Electric current and noise}

At the lowest order of the perturbative expansion in the tunneling amplitude $ t_{\mathbf{k},\alpha}$, the average current reads~\cite{Jauhobook,Mahanbook} 
\begin{equation}
\!\!\!\langle I_C\rangle\! =\! -2 \pi e \sum_{\alpha}  \left| t_{\alpha} \right|^2 \!\!\! \int^\infty_{-\infty}  \!\!\! \!\!\! d\epsilon \nu_L \left( \epsilon \right) \nu_D \left( \epsilon \right)
\left[ f_L \left(\epsilon \right) - f_R \left(\epsilon \right) \right],
\label{eq:avgcurrent}
\end{equation}
where $\nu_L$ and $\nu_D$ are the density of states (DoSs) of the left lead and the QD at the tunnel barrier, respectively; $f_L$ and $f_R$ are the corresponding Fermi distribution functions. Here, we assume the tunneling to be energy-independent, allowing us to simplify $t_{\mathbf{k},\alpha} = t_{\alpha}$.

Electric noise, defined as $\mathcal{S}_C\left(t-t^\prime\right)$$=\left\langle I_C\left(t\right) I_C\left(t'\right) \right\rangle$$-\left\langle I_C\left(t\right)\right\rangle\left\langle  I_C\left(t^\prime\right) \right\rangle$, or, alternatively, its Fourier transform, $\mathcal{S}_C\left(\omega\right)$, has been studied extensively. Although equilibrium zero-frequency noise $\mathcal{S}_C = \mathcal{S}_C \left( \omega = 0 \right)$$= \int d\tau \left[ \mathcal{S}_C (\tau) + \mathcal{S}_C (-\tau) \right]$$= 2 \int d\tau \mathcal{S}_C(\tau)$ can be related to conductance by the fluctuation-dissipation theorem and does not carry additional information, out-of-equilibrium zero-frequency noise can yield information on charge fluctuations in the mesoscopic system~\cite{shotnoise1,shotnoise2,shotnoise3}. The zero-frequency noise corresponds to the fluctuations of the tunneling charge current. It is proportional to the local DoSs and related to the Fermi distributions as 
\begin{eqnarray}
\mathcal{S}_C &=& 4 \pi  e^2 \sum_{\alpha}  \left| t_{\alpha} \right|^2 \int^\infty_{-\infty} d\epsilon  \nu_L\left( \epsilon \right) \nu_D\left( \epsilon \right)\nonumber\\
&&\times\left[f_L \left(\epsilon \right) + f_R \left(\epsilon \right) -2 f_L \left(\epsilon \right) f_R \left(\epsilon \right)\right].
\label{eq:zerofreqnoise}
\end{eqnarray}
In the equilibrium case, we find the relation between zero-frequency noise $\mathcal{S}_C^{eq}$ and electric conductance $G$ as expected $\mathcal{S}_C^{eq}=4TG$.

\subsection{Heat current and noise}

We similarly obtain the expression for the average heat current as
\begin{eqnarray}
\langle I_Q \rangle 
&=&-2 \pi  \sum_{\alpha}  \left| t_{\alpha} \right|^2 \!\!\!\int^\infty_{-\infty} \!\!\! d\epsilon \left( \epsilon - \mu_L \right) \nu_1 \left( \epsilon \right) \nu_2 \left( \epsilon \right)\nonumber\\
&&\times\left[ f_L \left(\epsilon \right) - f_R \left(\epsilon \right) \right].
\label{eq:avgheatcurrent}
\end{eqnarray}

In the same spirit as for the charge noise introduced earlier, in order to probe fluctuations of the heat current, one should investigate heat noise, which is defined as $\mathcal{S}_Q (t-t^\prime)=$$\left\langle I_Q\left(t\right)I_Q\left(t^\prime\right) \right\rangle$$-\left\langle I_Q\left(t\right)\right\rangle\left\langle I_Q\left(t^\prime\right)\right\rangle$ with $I_Q$ being the tunneling heat current operator. The zero-frequency noise corresponds to the fluctuations of the tunneling heat current, and reads
\begin{eqnarray}
\mathcal{S}_Q &=& 4 \pi  \sum_{\alpha}  \left| t_{\alpha} \right|^2\!\!\! \int^\infty_{-\infty} \!\!\!\! d\epsilon  \left( \epsilon - \mu_L \right)^2 \nu_L \left( \epsilon \right) \nu_D \left( \epsilon \right)\nonumber\\
&&\times\left[f_L \left(\epsilon \right) + f_R \left(\epsilon \right) -2 f_L \left(\epsilon \right) f_R \left(\epsilon \right) \right].
\label{eq:zerofreqheatnoise}
\end{eqnarray}
We find that the zero-frequency heat noise at equilibrium, $\mathcal{S}_Q^{eq}$, is related to the thermal coefficient $G_H$ through $\mathcal{S}_Q^{eq}=-4T^2G_H$.

\subsection{Mixed noise}

At this stage, it is also possible to investigate the cross-correlations between charge and heat currents through the mixed noise defined as $\mathcal{S}_M (t-t')$$=\left\langle I_C\left(t\right) I_Q\left(t'\right)\right\rangle$$-\left\langle I_C\left(t\right)  \right\rangle  \left\langle  I_Q\left(t'\right) \right\rangle$, where $I_C$ and $I_Q$ are the tunneling charge and heat current operators, respectively. Along the same lines as the previous calculations for the charge and heat noises, the zero-frequency mixed noise can be written as
\begin{eqnarray}
\mathcal{S}_M &=& 4\pi e\sum_{\alpha}\left|t_{\alpha}\right|^2\!\!\!\int^\infty_{-\infty} \!\!\!\!\! d\epsilon\left( \epsilon - \mu_L \right) \nu_L \left( \epsilon \right) \nu_D \left( \epsilon \right)\nonumber\\
&&\times\left[f_L \left(\epsilon \right) + f_R \left(\epsilon \right) -2 f_L \left(\epsilon \right) f_R \left(\epsilon \right)\right].
\label{eq:zerofreqmixednoise}
\end{eqnarray}
The relation between zero-frequency mixed noise in the equilibrium situation $\mathcal{S}_M^{eq}$ and thermoelectric coefficient $G_T$ is $\mathcal{S}_Q^{eq}=-4T^2G_T$.

The computation of the average currents and zero-frequency noises in Eqs.~\eqref{eq:avgcurrent} through \eqref{eq:zerofreqmixednoise}, requires the explicit form of the DoSs. Because the left lead is a noninteracting Fermi sea, we can replace its DoS by its energy-independent value $\nu_{L,0}$ at the Fermi energy, while the DoS of the QD at the weak link is discussed in the next section.

\section{Correlation function $K(\tau)$ and density of states $\nu_D(\epsilon)$}

For the convenience of the calculations, following Matveev-Andreev theory~\cite{MAprb}, we replace $d_\alpha\rightarrow d_\alpha \hat{F}$ in Eq.~\eqref{eq:tunnelH}, where $\hat{F}$ is the charge-lowering operator, which obeys the commutation relation $\left[\hat{F},\hat{n}\right]=\hat{F}$. One should note that this substitution does not affect the form of the Hamiltonian $H_L$. The DoS $\nu_D(\epsilon)$ of the QD at the weak barrier is modified by the electron-electron interactions in the dot as
\begin{equation}
\nu_D(\epsilon)=\nu_{D,0}T_R\cosh\left(\frac{\epsilon}{2 T_R}\right)\!\!\!\int^\infty_{-\infty}\!\!\frac{e^{i\epsilon t} K\left(\frac{1}{2T_R}+it\right)}{\cosh(\pi T_R t)}dt,
\label{DoSdef}
\end{equation}
where $\nu_{D,0}$ stands for the DoS of the QD which is no longer renormalized by the electron-electron interactions, while the correlation function $K\left(1/2 T_R +it\right)$ characterizes these interactions [$K(\tau)$ $=\langle T_{\tau}\hat{F}(\tau)\hat{F}^{\dagger}(0)\rangle$ ($T_\tau$ is the time-ordering operator, the imaginary time $\tau$ runs from $0$ to $\beta=1/T_R$)].

The time-ordered correlation function is computed through the functional integration $K(\tau)=Z(\tau)/Z(0)$. We introduce the charge and spin fields $\phi_{c,s}(x,t)=\left[\phi_\uparrow (x,t)\pm \phi_\downarrow (x,t)\right]/\sqrt{2}$. Rewriting the Hamiltonian $H_R$ in Eq.~\eqref{HR} in terms of these variables, we find that the correlation function $K(\tau)$ can be factorized into charge and spin components as $K(\tau)=K_c(\tau)K_s(\tau)$ where the main contribution to $K_c(\tau)$ can be obtained in the limit $|r|=0$ (no backscattering at the QPCs). Previous studies have shown that the thermoelectric properties of the system are controlled by charge and spin fluctuations at low temperatures $T\ll E_C$. However, the effect of small but finite $|r|$ on the charge modes is negligible due to the Coulomb blockade in the QD, while any small backscattering amplitude pins spin-mode fluctuations and dramatically changes their low-frequency dynamics. The correlation function $K_s(\tau)$ is computed nonperturbatively \cite{MAprb}, enabling us to extend the results beyond perturbation theory for small $|r|$. The nonperturbative treatment involves the refermionization procedure~\cite{MAprl,MAprb,thanh2010,thanh2015}. The spin-mode model is mapped onto a time-dependent resonant scattering model.

In the end, the correlation function for the 2CK model in the nonperturbative treatment, concerning the reflection amplitudes of the QPCs, is expressed as~\cite{MAprb}
\begin{align}
K (\tau)=&\frac{\pi T_R\Gamma}{\gamma E_C}\frac{1}{|\sin(\pi T_R\tau)|}\int_{-\infty}^{\infty}\frac{d\omega~e^{\omega\tau}}{\left(\omega^{2}+\Gamma^{2}\right)\left(1+e^{\omega/T_R}\right)} \nonumber\\
&-4|r|^2\frac{T_R}{E_C}\frac{\sin\left(2\pi N\right)}{|\sin(\pi T_R\tau)|}\ln\left(\frac{E_C}{T_R+\Gamma}\right)\nonumber\\
&\times\int_{-\infty}^{\infty}\!\!\! d\omega\frac{\omega e^{\omega\tau}}{\left(\omega^{2}+\Gamma^{2}\right)\left(1+e^{\omega/T_R}\right)},
\label{eq:Knonper}
\end{align}
where $\Gamma$ is the Kondo-resonance width in the vicinity of Coulomb peaks 
\begin{eqnarray}
\Gamma\left(N\right)=\frac{8\gamma E_C}{\pi^{2}}|r|^2\cos^{2}(\pi N),
\label{eq:Gam}
\end{eqnarray}
and $\gamma=e^{\textbf{C}}\approx1.78$ ($\textbf{C}\approx 0.577$ is Euler's constant). Following Ref.\onlinecite{thanh2010,thanhcom2024}, the Kondo resonance width at a Coulomb peak is always finite for any asymmetry in the two Kondo channels. However, at this stage, we only study the symmetric case. One should also notice that the two terms in Eq.~\eqref{eq:Knonper} determine the parity of the QD's DoS. Indeed, substituting Eq.~\eqref{eq:Knonper} back into Eq.~\eqref{DoSdef}, one has $\nu_D(\epsilon)=\nu_D^e(\epsilon)+\nu_D^o(\epsilon)$ with 
\begin{eqnarray}
\nu_D^e(\epsilon) \!&=&\! \frac{\nu_{D,0} T_R}{2 \gamma E_C}\cosh \left( \frac{\epsilon}{2 T_R} \right)\nonumber\\
&&\times\!\!\int^\infty_{-\infty}\!\!\! dx \frac{1}{\cosh \left( \frac{x}{2} \right)} \frac{p}{x^2 + p^2} \frac{x + \frac{\epsilon}{T_R}}{\sinh \left( \frac{x+\epsilon/T_R}{2}\right)},\nonumber
\\ 
\nu_D^o (\epsilon) \!&=&\! - \frac{2\nu_{D,0} T_R}{\pi E_C} \left| r \right|^2 \sin \left(2 \pi N \right) \nonumber\\
&&\!\!\!\!\times\log \left( \frac{E_C}{T_R (1+p)} \right)  \cosh \left( \frac{\epsilon}{2 T_R} \right)\nonumber \\
&& \!\!\!\!\times \!\!\int^\infty_{-\infty}\!\!\!\!\! dx \frac{1}{\cosh \left( \frac{x}{2} \right)} \frac{x}{x^2 + p^2}  \frac{x + \frac{\epsilon}{T_R}}{\sinh \left( \frac{x+\epsilon/T_R}{2}\right)},
\label{DoS}
\end{eqnarray}
where $p=\Gamma/T_R$. These DoS components, based on their parity properties with respect to energy, will determine the characteristics of thermoelectric coefficients and noises. 

\section{Main results: Out-of-equilibrium situations} 

In this section, we investigate both charge and heat transport. We compute voltage-driven electric and heat noises when a voltage bias is applied at the weak link, while temperature-driven electric and heat noises are calculated when the temperature bias is enforced. Mixed noise, the electric current and heat current correlation function, is examined in both cases \cite{comment}. We calculate these quantities using the first-order approximation of the series expansion, based on the voltage or temperature difference across the weak link. Therefore, the condition for these results to be valid is that the voltage is sufficiently small, i.e., $V\ll E_C$,
and the temperature difference is small enough, such that $\Delta T \ll T \ll E_C$.

\subsection{Voltage bias and noises}

We consider the generation of voltage-driven noise under a non-equilibrium condition, in which the two sides of the weak link have the same temperature $T_R = T_L = T$, but the chemical potentials satisfy $\mu_R-\mu_L=eV$. Expanding the combinations of Fermi functions entering the expressions for the current and the noise up to first order in $eV$, we are left with
\begin{align}
f_L \left(\epsilon \right) - f_R \left(\epsilon \right)= &-\frac{1}{2} \frac{1}{\cosh^2 \left( \frac{\epsilon}{2T}\right)} \frac{eV}{2T}\nonumber\\
& + O \left[ \left( \frac{eV}{T} \right)^2 \right],
\end{align}
and similarly
\begin{align}
&f_L \left(\epsilon \right) + f_R \left(\epsilon \right) -2 f_L \left(\epsilon \right) f_R \left(\epsilon \right) =\nonumber\\
& \quad \frac{1}{2} \frac{1}{\cosh^2 \left( \frac{\epsilon}{2T}\right)}\left[ 1 - \frac{eV}{2 T} \tanh \left( \frac{\epsilon}{2 T} \right) \right] + O \left[ \left( \frac{eV}{T} \right)^2 \right].
\end{align}
We point out that we have adopted the units $\hbar=k_B=1$ and introduced the characteristic conductance $
G_L = 2 \pi e^2 \nu_{L,0} \nu_{D,0} \sum_{\alpha}  \left| t_{\alpha} \right|^2$, 
which corresponds to the conductance of the barrier when the electrons in the dot are noninteracting. In the end, we obtain the following formula for the average electric current
\begin{equation}
\langle I_C^V \rangle\!=\! V\frac{G_L}{2\gamma}\frac{T}{E_C}\!f_G\left(\frac{\Gamma}{T}\right)\!, 
\label{IeV}    
\end{equation}
where~\cite{kiselev2025}
\begin{equation}
f_G\!\left(p\right)\!= p + \frac{\pi}{2} \!\left(\! 1-\frac{p^2}{\pi^2} \!\right)\! \psi^{(1)}\!\! \left( \frac{1}{2} +\frac{p}{2 \pi}\! \right)\!, 
\label{FG}    
\end{equation}
while the shot noise \cite{comment} contribution (we ignored the equilibrium thermal contribution) is given by
 \begin{align}
 \label{noiseeV}
\Delta \mathcal{S}_C^V =& V \frac{e G_L}{\pi}\frac{T}{E_C} \left| r \right|^2 \sin \left(2 \pi N \right) \log \left[ \frac{E_C}{T+\Gamma} \right] f_C^V\left(\frac{\Gamma}{T}\right), \\
f_C^V\left(p\right) =& \left[ 12 - \frac{8 p}{ \pi} \psi^{(1)} \!\left( \frac{1}{2} +\frac{p}{2 \pi} \right)\right.\nonumber\\
&\left.+ \left(1-\frac{p^2}{\pi^2} \right) \psi^{(2)} \left( \frac{1}{2} + \frac{p}{2 \pi} \right) \right],
\label{FCeV}
\end{align}
where $\psi^{(1)}(x)=\sum_{n=0}^\infty(x+n)^{-2}$ is the trigamma function, and $\psi^{(2)}(x)=\partial_x \psi^{(1)}(x)$ is the tetragamma function.

From Eq.~\eqref{IeV}, we can obtain the formula of electric conductance $G$~\cite{thanh2010}. It is interesting that the shot noise in Eq.~\eqref{noiseeV} is proportional to the thermoelectric coefficient $G_T$ (see formula (69) in Ref.~\onlinecite{MAprb}). This can be understood that both properties are affected by the scattering processes and energy transport in the system. The thermoelectric coefficient (and thermopower) involves energy transport, and fluctuations in this energy transport contribute to shot noise. The electric conductance [can be extracted from formula \eqref{eq:avgcurrent}], on the other hand, is primarily related to charge transport and does not directly account for these energy-dependent fluctuations that affect thermoelectric properties. Therefore, when both charge and energy transport are considered, we find that shot noise is related to the conductance $G$ at zeroth order, and to the thermoelectric coefficient $G_T$ at first order in the expansion with respect to the voltage bias $V$ (see also \cite{PaKi2025a, PaKi2025b}).

In studying the heat-related transport properties, we also focused only on the excess noises by subtracting the corresponding equilibrium contribution. Since there are contributions of order $O(eV)$ arising from both the term in $\epsilon-\mu_L$ and the one involving Fermi functions, $\Delta \mathcal{S}_Q$ and $\Delta \mathcal{S}_M$ contain two terms. The average heat current is written as 
\begin{eqnarray}
\!\!\!\!\!\!\!\!\langle I_Q^V \rangle\!\! &=&\!\!V\!\frac{G_L}{3 e \pi}\frac{T^2}{E_C} \left| r \right|^2 \sin \left(2 \pi N \right)\!\log\!\left[\! \frac{E_C}{T+\Gamma} \!\right ]\! f_{G_T}\!\!\left(\!\frac{\Gamma}{T}\!\right)\!,
\label{IQavr}
\end{eqnarray}
where
\begin{eqnarray}
\!\!\!\!\!\!\!\! f_{G_T}\!\!\left(p\right)\!\! &=&\!\! \frac{8 \pi^2}{3}\! - 2 p^2   +  \left( p^2 - \pi^2 \right)\! \frac{p}{\pi} \psi^{(1)}\!\left( \!\frac{1}{2} +\frac{p}{2 \pi} \!\right)\!,
\label{FIQ}
\end{eqnarray}
while the corresponding voltage-driven heat noise is
\begin{eqnarray}
\!\!\!\!\!\!\!\!\Delta \mathcal{S}_Q^V\!\! &=&\!\! V\! \frac{G_L}{3\pi e}\frac{T^3}{E_C} \left| r \right|^2 \!\!\sin \left(2 \pi N \right) \!\log\!\left[\! \frac{E_C}{T+\Gamma} \!\right ]\!\! f_Q^V\!\left(\!\frac{\Gamma}{T}\!\right)\!, \label{shotnoiseQ} 
\end{eqnarray}
with
\begin{eqnarray}
f_Q^V\left(p\right)\!\! &=&\!\!\frac{82 \pi^2}{3} - 14 p^2 - 8\pi p \left(\! 2-\frac{p^2}{\pi^2} \!\right) \!\psi^{(1)} \!\!\left(\! \frac{1}{2} +\frac{p}{2 \pi} \!\right) \nonumber\\
&&\!\!\!+\frac{1}{2}\left(3 \pi^2 - 4 p^2  + \frac{p^4}{\pi^2} \right)   \psi^{(2)} \left( \frac{1}{2} + \frac{p}{2 \pi} \right) \!, \label{FQ} 
\end{eqnarray}
and the voltage-driven mixed noise is expressed as
\begin{eqnarray}
\Delta \mathcal{S}_M^V\!\!\! &=&\! V\frac{G_L}{3\gamma}\frac{T^2}{E_C} f_M^V\!\left(\!\frac{\Gamma}{T}\!\right)\!,\label{shotnoiseM}
 \end{eqnarray}
where
\begin{eqnarray}
f_M^V\left(p\right)\! &=&\! 5p +\pi\!\left(\!1-\frac{3p^2}{\pi^2}\!\right)\! \psi^{(1)}\! \left( \!\frac{1}{2}\!+\!\frac{p}{2 \pi} \!\right)\nonumber\\
&&\!\!\!+\frac{p}{4}\left(1-\frac{p^2}{\pi^2}\right) \psi^{(2)} \left( \frac{1}{2} + \frac{p}{2 \pi} \right)\!\!,
 \end{eqnarray}
 
From Eqs.~\eqref{IQavr} and \eqref{shotnoiseQ}, we observe that the voltage-driven heat noise of the heat current is proportional to the thermoelectric coefficient $G_T$ because $G_T=\langle I_Q^V \rangle/V$ in the linear regime. This can be explained as follows: The thermoelectric coefficient dictates the coupling between charge and energy, affecting both the steady-state heat current and the fluctuations in energy transport. The average heat current, driven by the charge current and the Seebeck coefficient, establishes the scale for energy fluctuations (shot noise) in the system. Both the thermoelectric coefficient and the average heat current play essential roles in determining the magnitude of shot noise, as they influence the transport of charge and energy, as well as the fluctuations that arise in these processes within the heat current.

\begin{figure}[t]
\begin{center}
\includegraphics[scale=0.24]{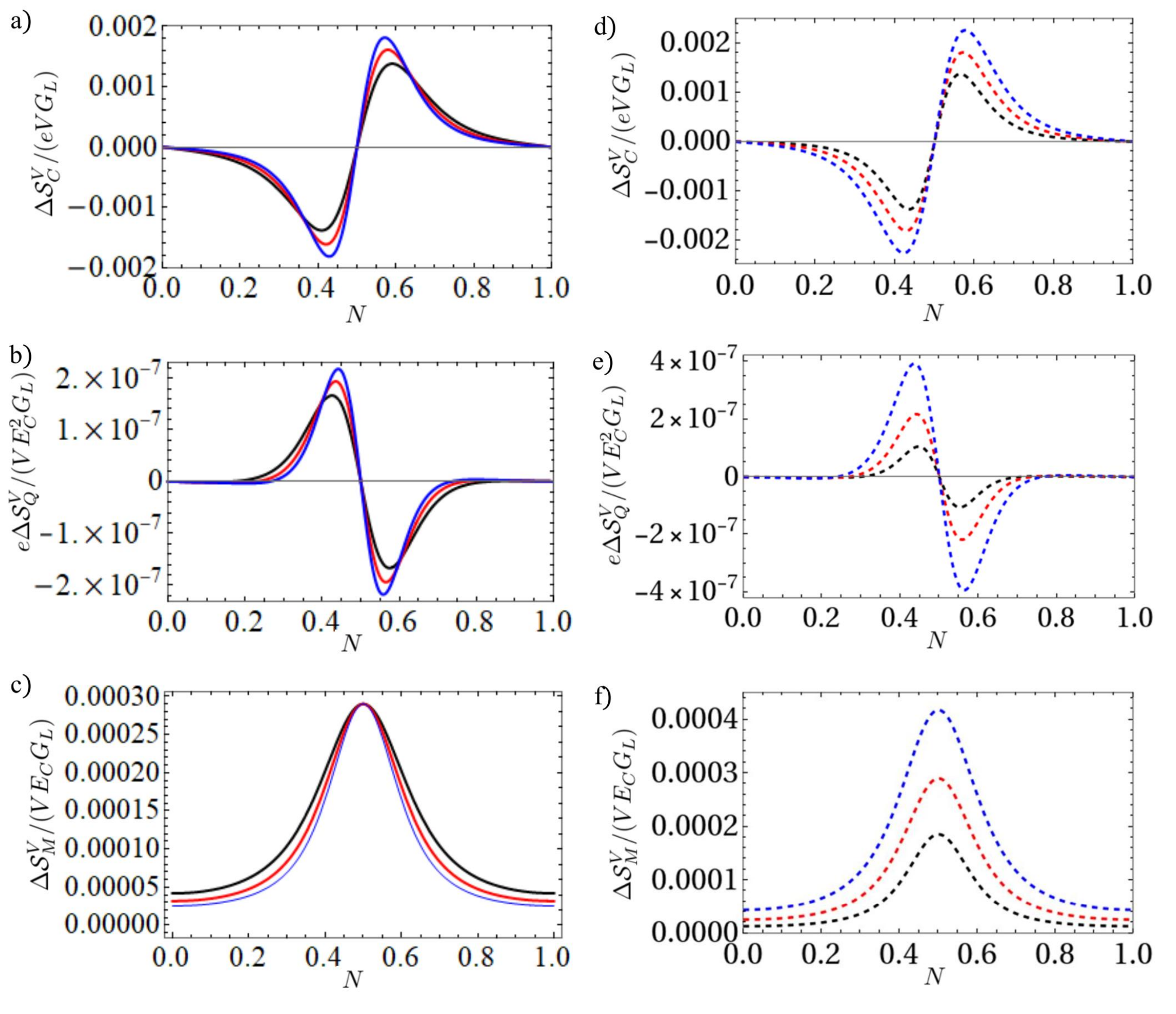}
\vspace{-0.3cm}
\caption{Voltage-driven electric noise $\Delta \mathcal{S}_C^V/(e G_L)$ [panels a) and d)], voltage-driven heat noise $e\Delta \mathcal{S}_Q^V/(E_C^2 G_L)$ [panels b) and e)], and voltage-driven mixed shot noise $\Delta \mathcal{S}_M^V/(E_C G_L)$ [panels c) and f)] over the voltage difference $V$ between two sides of the weak link as a function of the gate voltage $N$. For the plots on the left [a), b), and c)], $T/E_C=0.01$, black, red, and blue lines correspond to $|r|^2=0.06$, $|r|^2=0.08$, and $|r|^2=0.1$. For the plots on the right [d), e), and f)], $|r|^2=0.1$, black, red, and blue lines correspond to $T/E_C=0.008$, $T/E_C=0.01$, and $T/E_C=0.012$.}
\label{result1}
\end{center}
\end{figure} 

The gate voltage dependence of the voltage-driven noises $\Delta \mathcal{S}_C^V$, $\Delta \mathcal{S}_Q^V$, and $\Delta \mathcal{S}_M^V$ for both electric and heat currents, under the condition of a small voltage bias applied across the tunnel barrier, is shown in Fig.~\ref{result1}. While the voltage-driven noises $\Delta \mathcal{S}_C^V$ and $\Delta \mathcal{S}_Q^V$ of individual electric and heat currents are odd functions, the voltage-driven mixed noise $\Delta \mathcal{S}_M^V$ of these two currents is an even function of the gate voltage. Increasing the backscattering amplitude of the QPC leads to a slight elevation at the peaks in $\Delta \mathcal{S}_C^V$ and $\Delta \mathcal{S}_Q^V$, but no such effect is observed for $\Delta \mathcal{S}_M^V$. This is demonstrated in Fig.~\ref{result3} a), b), and c). 

\subsection{Temperature bias and noises}

We now turn to the situation of a temperature bias across the tunnel barrier. We set the temperatures $T_L = T + \Delta T$ and $T_R = T$, while enforcing the same chemical potentials $\mu_R = \mu_L$.  Expanding the combinations of Fermi functions entering the expressions for the current and the noise up to first order in $\Delta T$ (linear response regime), we are left with
\begin{align}
f_L \left(\epsilon \right) - f_R \left(\epsilon \right)=& \frac{1}{2} \frac{1}{\cosh^2 \left( \frac{\epsilon}{2T}\right)} \frac{\epsilon \Delta T}{2T^2}\nonumber\\
&+ O \left[ \left( \frac{\Delta T}{T} \right)^2 \right], 
\label{f1_deltaT}
 \end{align}
and
 \begin{align}
&f_L \left(\epsilon \right) + f_R \left(\epsilon \right) -2 f_L \left(\epsilon \right) f_R \left(\epsilon \right) =\nonumber\\
&\quad \frac{1}{2} \frac{1}{\cosh^2 \left( \frac{\epsilon}{2T}\right)} \left[ 1 + \frac{\epsilon \Delta T}{2 T^2} \tanh \left( \frac{\epsilon}{2 T} \right) \right]  \nonumber\\
&\quad + O \left[ \left( \frac{\Delta T}{T} \right)^2 \right] .
\label{f2_deltaT}
 \end{align}
Substituting the expression \eqref{f1_deltaT} into Eq.~\eqref{eq:avgcurrent} yields the formula for the average current 
\begin{eqnarray}
\langle I_C^{\Delta T} (t) \rangle\!\!& =&\! -\Delta T\frac{G_L}{3e\pi}\!\left| r \right|^2\!\sin \left(2 \pi N \right)\frac{T}{E_C}\!\log\!\left[\! \frac{E_C}{T+\Gamma} \!\right ]\nonumber\\
&&\!\times f_{G_T}\!\left(\!\frac{\Gamma}{T}\!\right)\!.
\label{Iavr_deltaT}
\end{eqnarray}
We notice that Eqs.~\eqref{IQavr} and \eqref{Iavr_deltaT}, together with the expressions for $G$ and $G_T$ in Ref.\cite{MAprb}, precisely satisfy the Onsager relation as presented in Eq.~\eqref{OnsagerRelation}. Meanwhile, plugging the expression \eqref{f2_deltaT} into Eq.~\eqref{eq:zerofreqnoise} gives the formula for the zero-frequency excess delta-T noise as follows:
\begin{eqnarray}
\Delta \mathcal{S}_C^{\Delta T}\!\!\!\! &=&\!\!\Delta T  \frac{G_L}{3 \gamma}\frac{T}{E_C} f_C^{\Delta T}\!\left(\!\frac{\Gamma}{T}\!\right)\!,
\label{deltaT_noise}
\end{eqnarray}
where
\begin{eqnarray}
 f_C^{\Delta T}\!\left(p\right)\! &=&\!\! p + 2 \pi \psi^{(1)} \left( \frac{1}{2} +\frac{p}{2 \pi} \right)\nonumber\\
&&\!\!\!\!\! -\frac{p}{4}\left(1-\frac{p^2}{\pi^2} \right) \psi^{(2)} \left( \frac{1}{2} + \frac{p}{2 \pi} \right)\!.
\label{deltaT_FC}
\end{eqnarray}
Concerning the fluctuations, together with the delta-T noise in Eq.~\eqref{deltaT_noise}, it is also important to investigate both the temperature-driven heat and mixed noises. By applying the zero-voltage and finite-temperature bias conditions into Eqs.~\eqref{eq:avgheatcurrent}, \eqref{eq:zerofreqheatnoise}, and \eqref{eq:zerofreqmixednoise}, we derive the formula for the average heat current as
\begin{eqnarray}
\!\!\! \langle I_Q^{\Delta T} \rangle\!\!\! &=&\!  - \Delta T  \frac{G_L}{48 e^2\gamma}\frac{T^2}{E_C} f_{G_H}\!\left(\!\frac{\Gamma}{T}\!\right)\!,
 \label{IQavr_deltaT}
\end{eqnarray}
where
\begin{eqnarray}
\!\!\!\!\!\! f_{G_H}\!\left(p\right)\!\! &=&\!\! 4 p \frac{13 \pi^2}{3}  - 4 p^3 \nonumber\\
&&\!\!\!\!\!\! + \left(3 \pi^4  -4\pi^2  p^2 + p^4 \right)\!\frac{2}{\pi}\psi^{(1)}\!\!\left(\!\frac{1}{2} +\frac{p}{2 \pi}\!\right)\!\!. 
 \label{deltaT_FGH}
\end{eqnarray}
The temperature-driven heat noise is obtained as
\begin{eqnarray}
\!\!\!\!\Delta \mathcal{S}_Q^{\Delta T}\!\!\!\! &=&\!\! \Delta T   \frac{G_L}{120 e^2 \gamma}\frac{T^3}{E_C} f_Q^{\Delta T}\!\left(\!\frac{\Gamma}{T}\!\right)\!,
\label{deltaT_heatnoise}
\end{eqnarray}
with
\begin{eqnarray}
\!\!\!\!\!\! f_Q^{\Delta T}\!\left(p\right)\!\! &=&\!\! 92 \pi^2 p -12 p^3 + 112 \pi^3  \psi^{(1)} \left(\frac{1}{2} +\frac{p}{2 \pi} \right)\nonumber\\
&&\!\!\!\!\!\! -p \!\left(\! 17\pi^2  -  20  p^2 + \frac{3 p^4}{\pi^2}\!\right)  \!\psi^{(2)} \!\!\left(\! \frac{1}{2} + \frac{p}{2 \pi}\!\right)\!\!.
\label{deltaT_FQ}
\end{eqnarray}
The temperature-driven mixed noise in this situation is written as 
\begin{eqnarray}
\!\!\Delta \mathcal{S}_M^{\Delta T} \!\!\!\! &=& \!\! - \Delta T  \frac{G_L}{6\pi e} \frac{T^2}{E_C}\left| r \right|^2 \sin \left(2 \pi N \right) \!\log \!\left[\!\frac{E_C}{T+\Gamma} \!\right]\nonumber\\
&&\!\!\!\times f_M^{\Delta T}\!\left(\!\frac{\Gamma}{T}\!\right)\!,
\label{deltaT_heatmixednoise}
\end{eqnarray}
where
\begin{eqnarray}
\!\!\!\!\!\! f_M^{\Delta T}\!\left(p\right)\!\! &=& \! 12\pi^2   - 16 \pi p \psi^{(1)} \left( \frac{1}{2} +\frac{p}{2 \pi} \right) + 4 p ^2 \nonumber\\
&&\!\!\! + \left(3 \pi^2  - 4 p^2 +  \frac{p^4}{\pi^2} \right)  \psi^{(2)} \left( \frac{1}{2} + \frac{p}{2 \pi} \right)\!\!.
\label{deltaT_FM}
\end{eqnarray}

\begin{figure}[t]
\begin{center}
\includegraphics[scale=0.24]{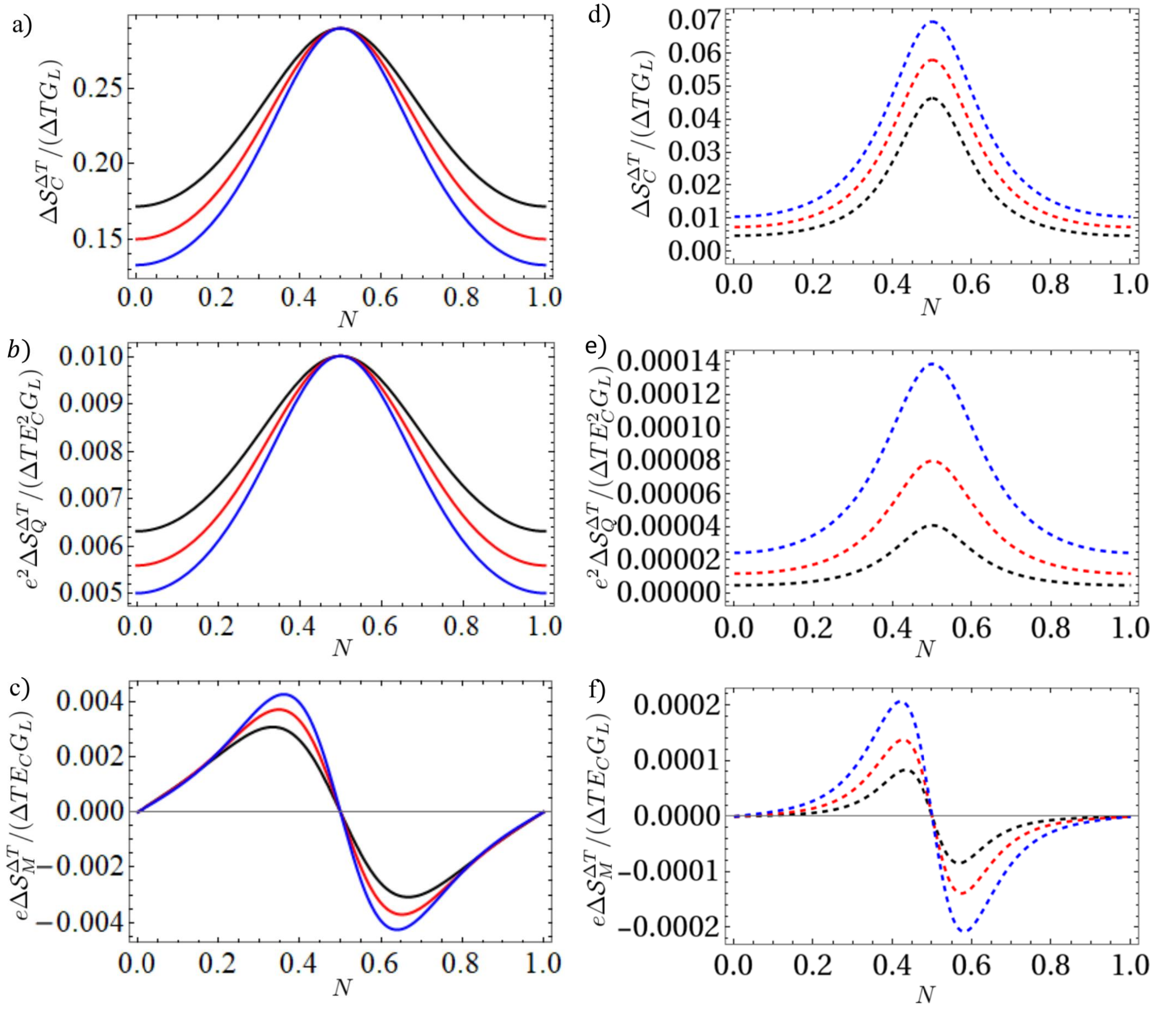}
\vspace{-0.3cm}
\caption{Temperature-driven charge noise (delta-T noise) $\Delta \mathcal{S}_C^{\Delta T}/G_L$ [panels a) and d)], temperature-driven heat noise $e^2\Delta \mathcal{S}_Q^{\Delta T}/(E_C^2G_L)$ [panels b) and e)], and temperature-driven mixed noise $e\Delta \mathcal{S}_M^{\Delta T}/(E_C G_L$ [panels c) and f)] over the temperature difference $\Delta T$ between two sides of the weak link $\Delta T/E_C$ as a function of the gate voltage $N$. For the plots on the left [a), b), and c)], $T/E_C=0.01$, black, red, and blue lines correspond to $|r|^2=0.06$, $|r|^2=0.08$, and $|r|^2=0.1$. For the plots on the right [d), e), and f)], $|r|^2=0.1$, black, red, and blue lines correspond to $T/E_C=0.008$, $T/E_C=0.01$, and $T/E_C=0.012$. }
\label{result2}
\end{center}
\end{figure} 

The temperature-driven electric and heat noises in Eqs.~\eqref{deltaT_noise} and \eqref{deltaT_heatnoise} are completely different from the voltage-driven electric and heat noises in Eqs.~\eqref{noiseeV} and \eqref{shotnoiseQ} for both electric and heat transport. The temperature-driven electric and heat noises are associated with the thermal coefficient $G_H$ (because $G_H=\langle I_Q^{\Delta T} \rangle/\Delta T$ in the linear regime), following the Wiedemann-Franz law~\cite{WF,kiselev2023}, they are also associated with the electric conductance $G$ (where $G=\langle I_C^{V} \rangle/V$ in the linear regime), while voltage-driven electric and heat noises are related to the electric coefficient $G_T$. (The expressions of $G, G_H, G_T$ can be seen easily in formulas of average currents.) Fig.~\ref{result2} shows the gate voltage dependence of temperature-driven elctric, heat, and mixed noises $\Delta \mathcal{S}_C^{\Delta T}$, $\Delta \mathcal{S}_Q^{\Delta T}$, and $\Delta \mathcal{S}_M^{\Delta T}$, in the presence of a temperature drop $\Delta T$ at the tunnel barrier. Temperature-driven noise and voltage-driven noise have the same partition origin but are activated by different stimuli, they thus behave completely different from each other.

\subsection{Symmetry properties}

The first two terms in the Hamiltonian describing the QD-QPC structure [see Eq.~\eqref{HR}] exhibit electron-hole (PH) symmetry. The third term, representing the charging energy, is invariant under the electron-hole transformation accompanied by the change of sign of the gate voltage: $N \rightarrow -N$. When $N$ is changed to $-N$, the Hamiltonian's form (without scattering at the QPCs) remains unchanged, provided the electrons are simultaneously converted into holes. However, the PH symmetry in this 2CK model is broken by scattering at the QPCs~\cite{MAprb}, leading to an imbalance in the thermal distributions of particles and holes. {\color{black} This imbalance explains why, under voltage bias, the heat current across the tunnel barrier is an odd function of the gate voltage $N$, while the electric current, which reflects charge transport rather than energy transport, does not exhibit the same symmetry. In contrast, when a temperature difference is applied, the roles reverse: it is the electric current that becomes an odd function of $N$, reflecting the asymmetry in energy transport.}

The voltage-driven noises and temperature-driven noises, which arise from random fluctuations in both heat and charge transport due to the voltage bias and temperature difference, reflect fluctuations in transport rather than the net flow of carriers. Therefore, the voltage-driven noises in Eqs.~\eqref{noiseeV} and \eqref{shotnoiseQ}, as well as the temperature-driven noises in Eqs.~\eqref{deltaT_noise} and \eqref{deltaT_heatnoise}, are unaffected by whether the system is PH symmetric or asymmetric. Namely, $\Delta \mathcal{S}_C^V$, $\Delta \mathcal{S}_Q^V$, and $\Delta \mathcal{S}_M^{\Delta T}$ vanish when the system is PH symmetric. The other noises remain finite.

Regarding the gate voltage dependence, the ratios $\Delta \mathcal{S}_C^V/V$, $\Delta \mathcal{S}_Q^V/V$, and $\Delta \mathcal{S}_M^{\Delta T}/\Delta T$ are odd functions of $N$. In contrast, the ratios $\Delta \mathcal{S}_M^V/V$, $\Delta \mathcal{S}_C^{\Delta T}/\Delta T$, and $\Delta \mathcal{S}_Q^{\Delta T}/\Delta T$ are even functions of $N$, as demonstrated in Figs.~\ref{result1} and \ref{result2}.  This can be understood by considering time-reversal symmetry (TRS). Under time reversal, the voltage bias $V$ changes sign, whereas the temperature difference $\Delta T$ remains unchanged. Although both the electric current $I_C$ and the heat current $I_Q$ are odd under TRS, they originate from different physical mechanisms. As a result, their cross-correlation $\langle I_C(t)I_Q(t^\prime)\rangle$ is odd under TRS, while the auto-correlations $\langle I_C(t)I_C(t^\prime)\rangle$ and $\langle I_Q(t)I_Q(t^\prime)\rangle$ are even under TRS.

\begin{figure}[t]
\begin{center}
\includegraphics[scale=0.24]{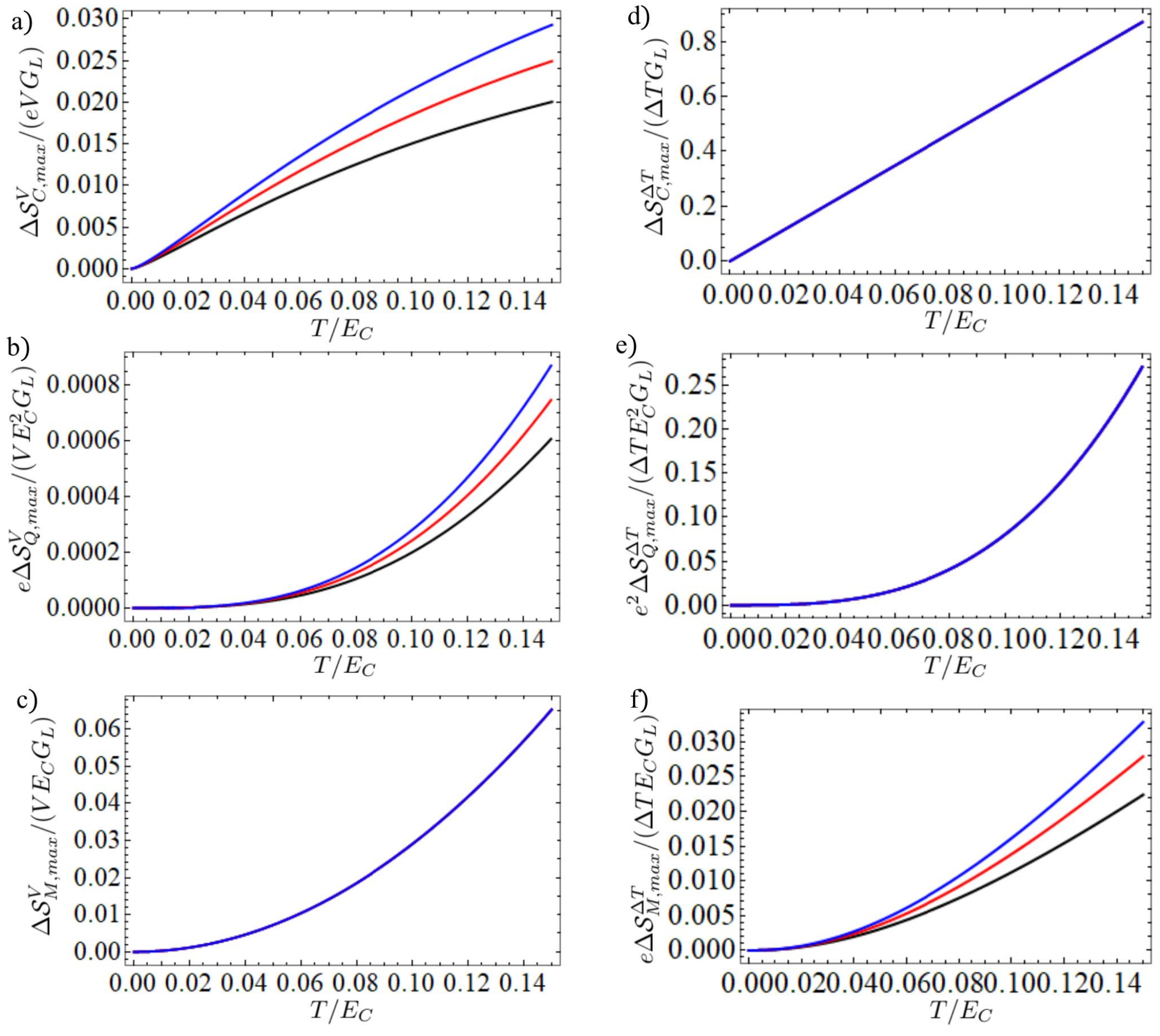}
\vspace{-0.3cm}
\caption{Maximum of voltage-driven electric noise (shot noise) $\Delta\mathcal{S}_{C,max}^V/(eG_L)$ [panel a)], maximum of voltage-driven heat noise $e\Delta\mathcal{S}_{Q,max}^V/(E_C^2G_L)$ [panel b)], maximum of voltage-driven mixed noise $\Delta\mathcal{S}_{M,max}^V/(E_CG_L)$ [panel c)] over the voltage difference $V$ between two sides of the weak link, maximum of temperature-driven electric noise $\Delta\mathcal{S}_{C,max}^{\Delta T}/(G_L)$ [panel d)], maximum of temperature-driven heat noise $e^2\Delta\mathcal{S}_{Q,max}^{\Delta T}/(E_C^2G_L)$ [panel e)], and maximum of temperature-driven mixed noise $e\Delta\mathcal{S}_{M,max}^{\Delta T}/(E_C G_L)$ [panel f)] over the temperature difference $\Delta T$ between two sides of the weak link as functions of temperature $T/E_C$ are plotted. For all plots,  black, red, and blue lines are corresponding to $|r|^2=0.06$, $|r|^2=0.08$, and $|r|^2=0.1$.}
\label{result3}
\end{center}
\end{figure} 

Figure \ref{result3} demonstrates the monotonic temperature dependence of the maxima of voltage-driven noise and temperature-driven noise for electric current, heat current, and their combination, for different values of reflection amplitudes at the QPC. When the noise expressions as a function of $N$ are even, the noises exhibit the same maximum value at $N=0.5$ for all $r$. However, when the noise expressions are odd, the maximum values increase with $r$. In terms of thermoelectric properties, we observe that as the reflection at the QPC increases, both the maximum of the thermopower [$S_{max}\sim |r|^2 \ln(|r|^2)]$ and the figure of merit [$ZT_{max}\sim |r|^4 \ln^2(|r|^2)]$ are enhanced \cite{MAprb,thanh2024,thanhcom2024}.  
Simultaneously, voltage-driven noise of both the electric current and heat current increases, while the temperature-driven noise remains unchanged. These findings are promising when investigating 2CK as a thermoelectric material.
\begin{figure}[t]
\begin{center}
\includegraphics[scale=0.25]{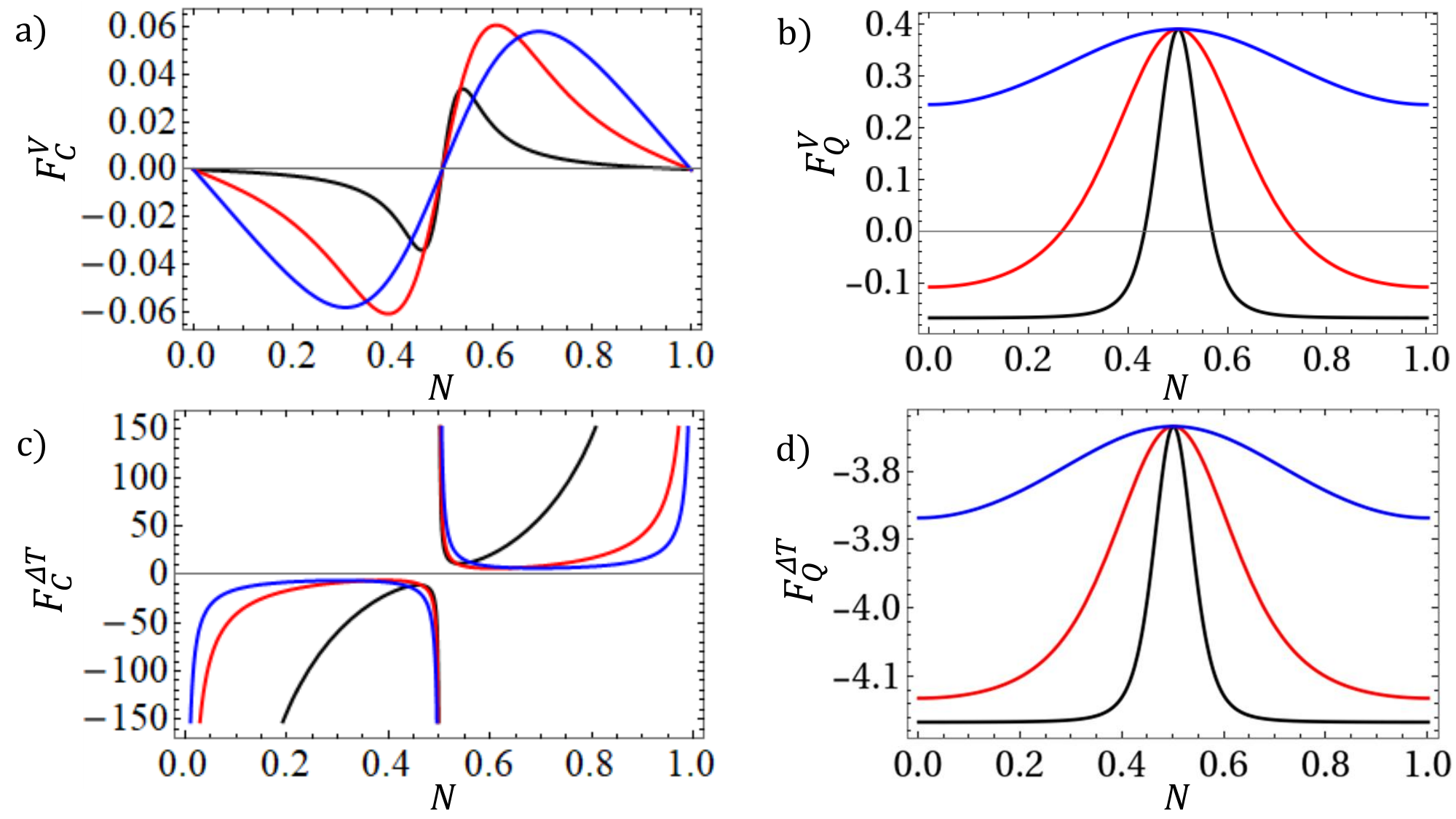}
\vspace{-0.3cm}
\caption{Fano factors as functions of the gate voltage $N$: a) $F_C^{V}$, b) $F_Q^{V}$, c) $F_C^{\Delta T}$, and d) $F_Q^{\Delta T}$ are plotted with different temperatures $T/E_C=0.001$, $T/E_C=0.01$, and $T/E_C=0.1$ corresponding to black, red, and blue lines. For all plots, $|r|^2=0.1$.}
\label{result4}
\end{center}
\end{figure} 

\subsection{Fano factors}

In this subsection, we examine the Fano factor, a dimensionless quantity that describes the statistical properties of fluctuations in particle counting. It is defined as $F_C=\Delta \mathcal{S}_C/2e\langle I_C\rangle$ for charge transport and $F_Q=\Delta \mathcal{S}_Q/2T\langle I_Q\rangle$ for heat transport. 

The Fano factors $F_C^{V}$ [panel a)], $F_Q^{V}$ [panel b)], $F_C^{\Delta T}$ [panel c)], and $F_Q^{\Delta T}$[panel d)] as a function of the gate voltage are plotted for different temperatures in Fig.~\ref{result4}. Interestingly, the Fano factors $F_C^{V}$ and $F_C^{\Delta T}$ either vanish or diverge in the Coulomb peaks, but their product is always finite, yielding
\begin{eqnarray}
&&\!\!\!\!\!\!\!\!\!\!\!\!\!\!\!\!\!\! F_C^{V} F_C^{\Delta T}\nonumber\\ 
&&\!\!\!\!\!\!\!\!\!\!\!\!\!\!\!\!\!\! = \! -\frac{
   12 - \frac{8 p}{ \pi} \psi^{(1)} \left( \frac{1}{2} +\frac{p}{2 \pi} \right) + \left(1-\frac{p^2}{\pi^2} \right) \psi^{(2)} \left( \frac{1}{2} + \frac{p}{2 \pi} \right) 
}{   p + \frac{\pi}{2} \left(1-\frac{p^2}{\pi^2} \right) \psi^{(1)} \left( \frac{1}{2} +\frac{p}{2 \pi} \right)  } \nonumber \\
&&\!\!\!\!\!\!\!\!\!\!\!\!\!\!\!\!\!\!\times 
\frac{  4 p +  8 \pi \psi^{(1)} \left( \frac{1}{2} +\frac{p}{2 \pi} \right)  - p \left(1-\frac{p^2}{\pi^2} \right) \psi^{(2)} \left( \frac{1}{2} + \frac{p}{2 \pi} \right)  }{ 
     16 \left[\frac{4 \pi^2}{3}  - 16 p^2 - 8\pi p \left(1-\frac{p^2}{\pi^2} \right)  \psi^{(1)} \left( \frac{1}{2} +\frac{p}{2 \pi} \right)\right]  
 } \! ,
\end{eqnarray}
which reduces, in the two limiting regimes, to
\begin{align}
F_C^{V} F_C^{\Delta T} & \overset{\Gamma \ll T}{\longrightarrow} -3 \frac{ 12  +  \psi^{(2)} \left( \frac{1}{2}  \right) }{   4 \pi^2  }\approx 0.37 , \label{productFClim1}\\
F_C^{V} F_C^{\Delta T} & \overset{\Gamma \gg T}{\longrightarrow} \frac{15 }{4 \pi^2} \approx 0.38 .\label{productFClim2}
\end{align}
In fact, by relating the Fano factors with the thermopower in the linear response regime $S=G_T/G$ of the system, we can make a conjecture as $F_C^{V}\times S=\text{constant}\times (T/E_C)^\alpha$, while $F_C^{\Delta T}/S=\text{constant}\times (T/E_C)^{-\alpha}$ with $\alpha >0$ is a constant [where for 2CK system, $\alpha=1/2$ in the non-perturbative (low temperature) regime and $\alpha=0$ in the perturbative (higher temperature) regime]. Therefore, the product $\left(F_C^{V} F_C^{\Delta T}\right)$ is a universal constant. 

The heat Fano factors $F_Q^{V}$ [Fig.~\ref{result4} b)], and $F_Q^{\Delta T}$[Fig.~\ref{result4} d)] reach their maximum/minimum absolute values in the vicinity of the Coulomb peaks. 
Their product is
\onecolumngrid
\begin{eqnarray}
F_Q^{V} F_Q^{\Delta T} &=& \! -\frac{
   41\pi^2 -21p^2- \frac{12 p}{ \pi}\left(2\pi^2-p^2\right) \psi^{(1)} \left( \frac{1}{2} +\frac{p}{2 \pi} \right) +\frac{3}{4} \left(3\pi^2-4p^2+\frac{p^4}{\pi^2} \right) \psi^{(2)} \left( \frac{1}{2} + \frac{p}{2 \pi} \right) 
}{ 8\pi^2-6 p^2 + \frac{3p}{\pi} \left(p^2-\pi^2 \right) \psi^{(1)} \left( \frac{1}{2} +\frac{p}{2 \pi} \right)  } \nonumber \\
&&\!\!\!\!\!\!\!\!\!\!\!\!\!\!\!\!\!\!\times 
\frac{  92\pi^2 p -12p^3+ 112\pi^3 \psi^{(1)} \left( \frac{1}{2} +\frac{p}{2 \pi} \right)  - p \left(17\pi^2- 20 p^2+\frac{3p^4}{\pi^2} \right) \psi^{(2)} \left( \frac{1}{2} + \frac{p}{2 \pi} \right)  }{ 
     5 \left[4p\frac{13 \pi^2}{3}  - 4 p^3 +\frac{2}{\pi} \left(3\pi^4 - 4\pi^2 p^2+p^4 \right)  \psi^{(1)} \left( \frac{1}{2} +\frac{p}{2 \pi} \right)\right] 
 } \! ,
\end{eqnarray}
\twocolumngrid
which also reduces, in the two limiting regimes, to
\begin{align}
F_Q^{V} F_Q^{\Delta T} & \overset{\Gamma \ll T}{\longrightarrow} -\frac{287}{15}+\frac{63}{60}\psi^{(2)} \left(\frac{1}{2}\right) \approx -1.4631 , \label{productFQlim1}\\
F_Q^{V} F_Q^{\Delta T} & \overset{\Gamma \gg T}{\longrightarrow} \frac{25}{36} \approx 0.6944 .\label{productFQlim2}
\end{align}
This universal behavior is in full agreement with the general theory establishing the universal relations between thermoelectrics and noise in mesoscopic transport across a tunnel junction \cite{PaKi2025a}.

From Fig.~\ref{result4}, we observe that near the Coulomb peaks, a voltage bias suppresses both charge and heat current fluctuations, as indicated by Fano factors less than one. In contrast, a temperature bias significantly enhances these fluctuations, resulting in Fano factors greater than one. Under a voltage bias, the charge Fano factor vanishes and changes sign at the Coulomb peak, while the heat Fano factor reaches its maximum at the same point. At low temperatures, the heat Fano factor can even approach zero and become negative in the vicinity of the Coulomb peak. When a temperature difference is applied, the charge Fano factor exhibits a discontinuous behavior at the Coulomb peak. Meanwhile, the heat Fano factor again attains its maximum at the peak but remains negative. In both cases, whether driven by voltage or temperature bias, the charge Fano factor displays antisymmetric behavior around the Coulomb peak, while the heat Fano factor remains symmetric.

\subsection{Comparison with one-channel Kondo model}

Previous studies \cite{MAprb,thanh2010} have shown that thermoelectric transport properties differ significantly between 1CK and 2CK systems. In the 1CK regime, the thermoelectric coefficients are consistent with FL behavior, whereas in the 2CK regime, they exhibit NFL characteristics. Therefore, in this subsection, we compare the results of the 2CK circuit with those of the 1CK case to gain a deeper understanding of the distinct physical signatures and emergent phenomena associated with multi-channel Kondo systems.

The charge Kondo model with a 1CK exhibits FL behavior and can be treated perturbatively~\cite{MAprb}. The DoS of the QD at the weak link is given by:
\begin{eqnarray}
&& \nu_D^{1CK}(\epsilon)=\frac{\nu_{D0}\pi^2}{2\gamma^2E_C^2}\left\{\left[1-2\gamma\xi|r|\cos\left(2\pi N\right)\right]\left(\epsilon^2+\pi^2T_R^2\right)\right.\nonumber\\
&&\left.+\frac{4\pi\gamma\xi}{3 E_C}|r|\sin\left(2\pi N\right)\epsilon\left(\epsilon^2+\pi^2T_R^2\right)\right\},
\end{eqnarray}
where $\xi$ is a numerical constant, $\xi\approx 1.59$. The perturbative solution is valid in the temperature regime $|r|^2E_C\ll T \ll E_C$, with the requirement $|r|\ll 1$.

\subsubsection{Voltage bias and noises}

The average charge current is:
\begin{equation}
\langle I_C^{V,1CK}\rangle=V\left[1-2\gamma\xi|r|\cos\left(2\pi N\right)\right]\frac{2\pi^4 G_L}{3\gamma^2}\frac{T^2}{E_C^2}.
\end{equation}
The equilibrium charge noise is $\mathcal{S}_{C,eq}^{1CK}=\left[1-2\gamma\xi|r|\cos\left(2\pi N\right)\right]8G_L\pi^4T^3/3\gamma^2E_C^2=4GT$. The shot noise (linear in $eV$) is:
\begin{equation}
\Delta \mathcal{S}_C^{V,1CK}=-V\frac{8e\pi^5\xi G_L}{3\gamma}|r|\left(\frac{T}{E_C}\right)^3\sin\left(2\pi N\right),
\end{equation}
with the corresponding Fano factor:
\begin{equation}
F_C^{V,1CK}=-2\pi \xi\gamma\frac{T}{E_C}|r|\sin\left(2\pi N\right).
\label{FCV1CK}
\end{equation}

To compare with the 2CK case in the same temperature regime $|r|^2E_C\ll T \ll E_C$ (i.e. $\Gamma \ll T \ll E_C$), we write the explicit formula for the shot noise as shown in Eq.~\eqref{noiseeV} as:
\begin{equation}
\Delta \mathcal{S}_C^{V,2CK}\sim -V |r|^2\frac{T}{E_C}\ln\left(\frac{E_C}{T}\right)\sin\left(2\pi N\right).
\end{equation}
The charge Fano factor for 2CK in this case is
\begin{eqnarray}
F_C^{V,2CK}&\approx & \frac{4\gamma}{\pi^4}\left[12+\Psi^{(2)}\left(\frac{1}{2}\right)\right]\nonumber\\
&&\times|r|^2\ln\left(\frac{E_C}{T}\right)\sin\left(2\pi N\right).
\label{FCV2CK}
\end{eqnarray}

Similarly, the heat current of the 1CK setup is 
\begin{equation}
\langle I_Q^{V,1CK}\rangle=V\frac{8\pi^7\xi G_L}{15e\gamma}\frac{T^4}{E_C^3}|r|\sin\left(2\pi N\right).
\end{equation}
The equilibrium noise $\mathcal{S}_{Q,eq}^{1CK}=\left[1-2\gamma\xi|r|\cos\left(2\pi N\right)\right]8G_L\pi^6T^5/5\gamma^2E_C^2=4T^2 G_H $, and the voltage-induced part reads
\begin{equation}
\Delta \mathcal{S}_Q^{V,1CK}=-V\frac{8\pi^7\xi G_L}{45 e\gamma}\frac{T^5}{E_C^3}|r|\sin\left(2\pi N\right).
\end{equation}
The Fano factor is a constant $F_Q^{V,1CK}=-1/6\approx -0.1667$.

For 2CK in this condition, from Eq.~\eqref{shotnoiseQ}, we find the voltage-induced part as
\begin{equation}
\Delta \mathcal{S}_Q^{V,2CK}\sim V |r|^2\frac{T^3}{E_C}\ln\left(\frac{E_C}{T}\right)\sin\left(2\pi N\right),
\end{equation}
and the corresponding Fano factor $F_Q^{V,2CK}=41/8+9\psi^{(2)}(1/2)/32\approx 0.39$.

The mixed noise is a bit different from the above two noises. It includes the equilibrium term $\mathcal{S}_{M,eq}^{1CK}=\left(32G_L\pi^7\xi T^5/15\gamma E_C^3\right)|r|\sin\left(2\pi N\right)=-4G_T T^2$, and the voltage-induced component
\begin{equation}
\Delta \mathcal{S}_M^{V,1CK}=V\left[1-2\gamma\xi|r|\cos\left(2\pi N\right)\right]\frac{2\pi^4 G_L}{3\gamma^2}\frac{T^3}{E_C^2},
\end{equation}
while the mixed noise of 2CK in this temperature regime is obtained from Eq.~\eqref{shotnoiseM} as
\begin{equation}
\Delta \mathcal{S}_M^{V,2CK}\sim V |r|^2\frac{T^2}{E_C}.
\end{equation}

\subsubsection{Temperature bias and noises}

The average of the electric current in this situation is
\begin{equation}
\langle I_C^{\Delta T,1CK}\rangle=-\Delta T\frac{8\pi^7\xi G_L}{15e\gamma}\left(\frac{T}{E_C}\right)^3|r|\sin\left(2\pi N\right).
\end{equation}
The equilibrium delta-T noise $\mathcal{S}_{C,eq}^{\Delta T,1CK}=\left[1-2\gamma\xi|r|\cos\left(2\pi N\right)\right]8G_L\pi^4T^3/3\gamma^2E_C^2=4GT$, and the $\Delta T$-induced part is
\begin{equation}
\Delta \mathcal{S}_C^{\Delta T,1CK}=\Delta T\frac{2\pi^4 G_L}{\gamma^2}\left(\frac{T}{E_C}\right)^2\left[1-2\gamma\xi|r|\cos\left(2\pi N\right)\right].
\end{equation}
The Fano factor $F_C^{\Delta T,1CK}$ is
\begin{equation}
F_C^{\Delta T,1CK}\approx -\frac{15}{8\pi^3\gamma \xi}\frac{E_C}{T}\frac{1}{|r|\sin\left(2\pi N\right)}.
\label{FCT1CK}
\end{equation}
From Eqs.~\eqref{FCV1CK} and \eqref{FCT1CK}, we find that 
\begin{equation}
F_C^{V,1CK}F_C^{\Delta T,1CK}\approx\frac{15}{4\pi^2}\approx 0.38.
\label{productFC1CK}
\end{equation}
This is again in full agreement with the general theory \cite{PaKi2025a}.

For 2CK, from Eq.~\eqref{deltaT_noise}, we get (in the relevant regime $\Gamma \ll T$)
\begin{equation}
\Delta \mathcal{S}_C^{\Delta T,2CK}\sim \Delta T \left\{\frac{T}{E_C} + |r|^2\left[1+\cos\left(2\pi N\right)\right]\right\},
\end{equation}
and the Fano factor is
\begin{equation}
F_C^{\Delta T,2CK}\approx- \frac{3\pi^2}{16|r|^2\sin\left(2\pi N\right)\ln\left(\frac{E_C}{T}\right)}.
\end{equation}
We also re-obtain the product of Fano factors, which has been shown in Eq.~\eqref{productFClim1}. 
Interestingly, the product of the Fano factors for charge transport under voltage and temperature bias in both 1CK and 2CK models are numbers. 

For heat transport, the average current is
\begin{equation}
\langle I_Q^{\Delta T,1CK}\rangle=-\Delta T\frac{2G_L\pi^6}{5e^2\gamma^2}\frac{T^3}{E_C^2}\left[1-2\gamma\xi|r|\cos\left(2\pi N\right)\right].
\end{equation}
The temperature-driven heat noise includes $\mathcal{S}_{Q,eq}^{\Delta T,1CK}=\left[1-2\gamma\xi|r|\cos\left(2\pi N\right)\right]8G_L\pi^6T^5/5\gamma^2E_C^2=-4T^2G_H$ and 
\begin{equation}
\Delta \mathcal{S}_Q^{\Delta T,1CK}=\Delta T\frac{10 G_L\pi^6}{e^2 3\gamma^2}\frac{T^4}{E_C^2}\left[1-2\gamma\xi|r|\cos\left(2\pi N\right)\right].
\end{equation}
Analogous to the voltage-driven scenario, the heat Fano factor in response to a temperature bias is also found to be constant, $F_Q^{\Delta T,1CK}=-25/6\approx -4.1667$.

For 2CK, from Eq.~\eqref{deltaT_heatnoise}, we find
\begin{equation}
\Delta \mathcal{S}_Q^{\Delta T,2CK}\sim  \Delta T\left\{\frac{T^3}{E_C} + T^2|r|^2\left[1+\cos\left(2\pi N\right)\right]\right\},
\end{equation}
and the Fano factor $F_Q^{\Delta T,2CK}=-56/15\approx -3.73$. The product of the heat Fano factors in the high temperature regime of 1CK is c.f. \cite{PaKi2025a}
\begin{equation}
F_Q^{V,1CK}F_Q^{\Delta T,1CK}=\frac{25}{36}\approx 0.6944 .
\label{productFQ1CK}
\end{equation}
The dependence of the Fano factors on system parameters shows similar characteristics in both 1CK and 2CK systems.

Besides the equilibrium term $\mathcal{S}_{M,eq}^{\Delta T,1CK}=\left(32G_L\pi^7\xi T^5/15\gamma E_C^3\right)|r|\sin\left(2\pi N\right)=-4G_T T^2$, the temperature-bias-induced contribution to the mixed noise in the 1CK model is given by
\begin{equation}
\Delta \mathcal{S}_M^{\Delta T,1CK}=\Delta T\frac{G_L}{e}\frac{40\pi^7\xi}{9\gamma}\frac{T^4}{E_C^3}|r|\sin\left(2\pi N\right).
\end{equation}
The corresponding contribution for the 2CK model can be evaluated from Eq.~\eqref{deltaT_heatmixednoise} as
\begin{equation}
\Delta \mathcal{S}_M^{\Delta T,2CK}\sim -\Delta T |r|^2\frac{T^2}{E_C}\ln\left(\frac{E_C}{T}\right)\sin\left(2\pi N\right).
\end{equation}

We find that the symmetry of both the noise and the Fano factor with respect to the gate voltage $N$ is identical in the 1CK and 2CK models. Moreover, shot noise and delta-T noise of the charge current exhibit the behavior distinct from that of the corresponding differential electric conductance, much like the thermoelectric coefficient. In contrast, the mixed noise closely follows the behavior of the differential conductance. 

For the heat current, the voltage-driven noise and temperature-driven noise behave similarly to the corresponding differential heat conductance. However, this correspondence does not hold for the mixed noise, which deviates from the behavior of the differential heat conductance.

Despite these similarities, the temperature dependence of the noise differs significantly between the two models. In the 2CK case, logarithmic terms of the form $\ln(E_C/T)$ appear in the expressions for charge shot noise, heat voltage-driven noise, and mixed temperature-driven noise -- hallmarks of NFL behavior. These logarithmic contributions are absent in the 1CK case, which instead displays conventional FL behavior. The same distinction applies to the Fano factor.

Regarding PH symmetry, both the charge and heat shot noise vanish when PH symmetry is broken. Interestingly, the mixed shot noise in 2CK also vanishes under broken PH symmetry, whereas in 1CK it remains finite even in the PH-symmetric case.

For temperature-driven noise, we observe consistent behavior between the 1CK and 2CK models: the charge and heat temperature-driven noises both survive under PH symmetry, while the mixed delta-T noise vanishes when PH symmetry is broken.

From these observations, we propose that the measurement of mixed voltage-driven noise could serve as a sensitive probe to distinguish between FL and NFL behavior, particularly in relation to PH symmetry breaking.

The Fano factors of the heat currents in both cases (voltage and temperature bias) are constant in both charge Kondo models because both the heat current and its noise scale similarly in the low-temperature limit ($T\ll E_C$), governed by universal scattering processes around the Kondo fixed point. This leads to a universal constant ratio, a hallmark of the underlying quantum critical behavior in 2CK and FL behavior in 1CK.

\subsection{General reciprocity relations for noises}

 In this subsection, we briefly recapitulate the universal reciprocity relations established in \cite{PaKi2025a} for the general case of mesoscopic transport across a tunnel junction and illustrated by the theory for the noise signatures of a charged Sachdev-Ye-Kitaev dot \cite{PaKi2025b}.

The general expressions for the charge, heat, and mixed noise in the linear response regime can be derived for both the 1CK and 2CK charge Kondo models as follows:
\begin{eqnarray}
\mathcal{S}_C \!\!&=&\!\! 4TG - V\frac{eG_L}{2T}\mathcal{N}_0 + \Delta T\frac{G_L}{2T^2}\mathcal{N}_1, \nonumber\\ 
\mathcal{S}_Q \!\!&=&\!\! -4T^2G_H - V\!\left[8T^2G_T+\frac{G_L}{2eT}\mathcal{N}_2\!\right]\!\! + \!\Delta T\frac{G_L}{2e^2T^2}\mathcal{N}_3, \nonumber\\ 
\mathcal{S}_M \!\!&=&\!\! -4T^2G_T - V\!\left[-4TG+\frac{G_L}{2T}\mathcal{N}_1\!\right]\!\! + \!\Delta T\frac{G_L}{2eT^2}\mathcal{N}_2,
\end{eqnarray}
where for $l=0,1,2,3$ 
\begin{equation}
\mathcal{N}_l=\int_{-\infty}^{\infty}\epsilon^l\frac{\nu_D(\epsilon)}{\nu_{D,0}}\frac{\sinh\left(\frac{\epsilon}{2T}\right)}{\cosh^3\left(\frac{\epsilon}{2T}\right)} d\epsilon.    
\end{equation}

The ratios of the thermal (equilibrium) noises to the Onsager coefficients satisfy the so-called Onsager relations (Fluctuation-Dissipation theorem (FDT)). 
\begin{equation}
\frac{\mathcal{S}_C^{eq}}{G}=-\frac{1}{T}\frac{\mathcal{S}_Q^{eq}}{G_H}=-\frac{1}{T}\frac{\mathcal{S}_M^{eq}}{G_T}=4T.
\end{equation}
These ratios exhibit universal behavior. 
In particular, they are entirely independent of system parameters (for the 1CK and 2CK cases, these parameters are the number of channels, reflection amplitudes, etc.) and are determined solely by the temperature. Consequently, it can be straightforwardly shown that the Wiedemann-Franz law can be re-expressed in terms of the thermal noises, and the corresponding Lorenz ratio satisfies
\begin{equation}
R=-\frac{3}{\pi^2T^2}\left[\frac{\mathcal{S}_Q^{eq}}{\mathcal{S}_C^{eq}}+\left(\frac{\mathcal{S}_M^{eq}}{\mathcal{S}_C^{eq}}\right)^2\right].
\label{WFnoises}
\end{equation}

In the out-of-equilibrium linear response regime,
the mixed, voltage and temperature driven noises are not independent \cite{PaKi2025a, PaKi2025b}.
The general reciprocity relations for the noises are expressed as follows (see more details of the derivation in \cite{PaKi2025a, PaKi2025b}):
\begin{eqnarray}
\frac{\partial\mathcal{S}_M}{\partial V}+T\frac{\partial\mathcal{S}_C}{\partial \Delta T}&=&4TG,\label{RR1}\\
\frac{\partial\mathcal{S}_Q}{\partial V}+T\frac{\partial\mathcal{S}_M}{\partial \Delta T}&=&-8T^2G_T.
\label{RR2}
\end{eqnarray}

Equations (\ref{RR1}) and (\ref{RR2}) are exactly satisfied for Kondo models with an arbitrary number of channels. The validity of these reciprocal relations is not restricted by the specific model or by whether the system exhibits FL or NFL behavior. Moreover, these equations establish not only universal reciprocity relations among the three types of noise but also a universal connection between nonequilibrium noise and thermoelectric transport.

\section{Conclusions}

In summary, we have presented a detailed analysis of both electric and heat currents, as well as electric, heat, and mixed noises under small voltage and temperature biases. These noise characteristics exhibit distinct NFL behavior in the 2CK model, in contrast to the conventional FL behavior seen in the 1CK case. Importantly, noise signals are closely related to the thermoelectric coefficients in the linear response regime, establishing a direct link between fluctuations and transport properties.

We highlight unique features of exotic charge Kondo systems that are now accessible in experiments. By combining measurements of the thermoelectric coefficients and noises, we have provided valuable insights into the interplay between charge and heat transport in strongly correlated mesoscopic systems. The universality observed in the Fano factor products for electric and heat currents under both voltage and temperature biases reflects fundamental aspects of reversible thermoelectric effects and fixed-point physics.

The logarithmic temperature dependence characteristic of the 2CK model is further manifested in the thermopower, suggesting that nonlinear Seebeck coefficient and noise measurements offer sensitive probes of electron correlations in nanostructures. Although shot noise (voltage-driven electric noise) has been well studied in many mesoscopic and nanoscopic systems, both theoretically and experimentally, our focus on voltage-driven electric/heat noise,  voltage-driven mixed noise, and temperature-driven noises in the NFL regime of the two-channel charge Kondo quantum simulators uncovers complementary information about temperature-induced fluctuations and nonequilibrium dynamics. Our analysis also illustrates the new nonequilibrium reciprocity relations and a thermal-noise analogue of the Wiedemann-Franz law recently reported in \cite{PaKi2025a, PaKi2025b}, demonstrating a profound universality governed solely by temperature, independent of system-specific details. 

Overall, this work advances the fundamental understanding of quantum transport and noises in strongly correlated charge Kondo circuits, offering experimentally testable predictions that can guide future studies and the development of nanoscale thermoelectric devices operating beyond linear response.

\vspace{0.5cm}

\section*{Acknowledgements}
M.N.K  thanks Andrei Pavlov for numerous fruitful discussions and careful reading of the manuscript resulting in many useful suggestions. We thank T. Jonckheere for stimulating discussions. This research in Hanoi is funded by the Vietnam National Foundation for Science and Technology Development (NAFOSTED) under grant number 103.01-2023.03. T.K.T.N. acknowledges support from the ICTP through the Associates Programme (2024-2029). The work of M.N.K. is conducted within the framework of the Trieste Institute for Theoretical Quantum Technologies (TQT). T.K.T.N. and J.R. are co-first authors.


\begin{thebibliography}{100}

\bibitem{Wood} C. Wood, \textit{Materials for thermoelectric energy conversion}, Rep. Prog. Phys. \textbf{51}, 459 (1988).

\bibitem{TEbook1} D. K. C. MacDonald, \textit{Thermoelectricity: An Introduction to the Principles} (Wiley, New York, 1962).

\bibitem{TEbook2} H. J. Goldsmid, \textit{Introduction to Thermoelectricity}, (Springer-Verlag, Berlin Heidelberg, 2009).

\bibitem{TE_nano} A. J. Minnich, M. S. Dresselhaus, Z. F. Ren, G. Chen, Energy Environ. Sci. \textbf{2}, 466 (2009);  M. G. Kanatzidis, Chem. Mater. \textbf{22}, 648 (2009); C. J. Vineis, A. Shakouri, A. Majumdar, M. G. Kanatzidis, Adv. Mater. \textbf{22}, 3970 (2010); P. Vaqueiro, A. V. Powell, J. Mater. Chem. \textbf{20}, 9577 (2010); S. K. Bux, J. P. Fleurial, R. B. Kaner, Chem. Comm. \textbf{46}, 8311 (2010).

\bibitem{Blanterbook} Y. M. Blanter and Y. V. Nazarov, \textit{Quantum Transport: Introduction to Nanoscience} (Cambridge University Press, Cambridge, England, 2009).

\bibitem{Kiselevbook} K. Kikoin, M. N. Kiselev, and Y. Avishai, \textit{Dynamical Symmetry for Nanostructures. Implicit Symmetry in Single-Electron Transport Through Real and Artificial Molecules} (Springer, New York, 2012).

\bibitem{Kondo} J. Kondo, \textit{Resistance minimum in dilute magnetic alloys}, Prog. Theor. Phys. \textbf{32}, 37 (1964).

\bibitem{Hewson} A. Hewson, \textit{The Kondo Problem to Heavy Fermions} (Cambridge University Press, Cambridge, England, 1993).

\bibitem{Kondoreview} L. Kouwenhoven and L. Glazman, Phys. World \textbf{14}, 33 (2001); L. I. Glazman and M. Pustilnik, in \textit{Nanophysics: Coherence and Transport}, pp. 427, edited by H. Bouchiat \textit{et al.} (Elsevier, New York, 2005).

\bibitem{Nozieres} P. Nozi\'eres, \textit{A ``fermi-liquid" description of the Kondo problem at low temperatures}, J. Low Temp. Phys. \textbf{17}, 31 (1974). 

\bibitem{NozieresBlandin} P. Nozi\`eres and A. Blandin, \textit{Kondo effect in real metals}, J. Phys. France \textbf{41}, 193 (1980).

\bibitem{thanh2010} T. K. T. Nguyen, M. N. Kiselev, and V. E. Kravtsov, \textit{Thermoelectric transport through a quantum dot: Effects of asymmetry in Kondo channels}, Phys. Rev. B \textbf{82}, 113306 (2010).

\bibitem{unKondo1} S. Yasui and K. Sudoh, \textit{Heavy-quark dynamics for charm and bottom flavor on the Fermi surface at zero temperature}, Phys. Rev. C \textbf{88}, 015201 (2013); K. Hattori, K. Itakura, S. Ozaki, and S. Yasui, \textit{QCD Kondo effect: Quark matter with heavy-flavor impurities}, Phys. Rev. D \textbf{92}, 065003 (2015).

\bibitem{unKondo2} S. Yasui and K. Sudoh, \textit{Kondo effect of and mesons in nuclear matter}, Phys. Rev. C \textbf{95}, 035204 (2017); S. Yasui, \textit{Kondo effect in charm and bottom nuclei}, Phys. Rev. C \textbf{93}, 065204 (2016); S. Yasui and T. Miyamoto, \textit{Spin-isospin Kondo effects for $\Sigma_c$ and $\Sigma_c^*$ baryons and $\bar{D}$ and $\bar{D}^*$ mesons}, Phys. Rev. C \textbf{100}, 045201 (2019).

\bibitem{unKondo3} Y. Nishida, \textit{SU(3) Orbital Kondo effect with ultracold atoms}, Phys. Rev. Lett. \textbf{111}, 135301 (2013).

\bibitem{unKondo4} B. B\'eri and N. R. Cooper, \textit{Topological Kondo effect with Majorana fermions}, Phys. Rev. Lett. \textbf{109}, 156803 (2012).

\bibitem{flensberg} K. Flensberg, \textit{Capacitance and conductance of mesoscopic systems connected by quantum point contacts}, Phys. Rev. B \textbf{48}, 11156 (1993).

\bibitem{matveev} K. A. Matveev, \textit{Coulomb blockade at almost perfect transmission}, Phys. Rev. B \textbf{51}, 1743 (1995). 

\bibitem{furusakimatveev} A. Furusaki, K. A. Matveev, \textit{Theory of strong inelastic cotunneling}, Phys. Rev. B \textbf{52}, 16676 (1995). 

\bibitem{pierre2} Z. Iftikhar, S. Jezouin, A. Anthore, U. Gennser, F. D. Parmentier, A. Cavanna and F. Pierre, \textit{Two-channel Kondo effect and renormalization flow with macroscopic quantum charge states}, Nature \textbf{526}, 233 (2015).

\bibitem{pierre3} Z. Iftikhar, A. Anthore, A. K. Mitchell, F. D. Parmentier, U. Gennser, A. Ouerghi, A. Cavanna, C. Mora, P. Simon, and F. Pierre, \textit{Tunable quantum criticality and super-ballistic transport in a \textquotedblleft charge\textquotedblright{} Kondo circuit}, Science \textbf{360}, 1315 (2018).

\bibitem{michellprl} A. K. Mitchell, L. A. Landau, L. Fritz, and E. Sela, \textit{Universality and scaling in a charge two-channel Kondo device}, Phys. Rev. Lett. \textbf{116}, 157202 (2016).

\bibitem{thanhprl}  T. K. T. Nguyen and M. N. Kiselev, \textit{Thermoelectric transport in a three-channel charge Kondo circuit} Phys. Rev. Lett. \textbf{125}, 026801 (2020). 

\bibitem{2SCKC} W. Pouse, L. Peeters, C. L. Hsueh, U. Gennser, A. Cavanna, M. A. Kastner, A. K. Mitchell, and D. Goldhaber-Gordon, \textit{Quantum simulation of an exotic quantum critical point in a two-site charge Kondo circuit}, Nat. Phys. \textbf{19}, 492 (2023).

\bibitem{Beenakker}  C. W. J. Beenakker and A. A. M. Staring, \textit{Theory of the thermopower of a quantum dot}, Phys. Rev. B \textbf{46}, 9667 (1992).

\bibitem{Turek} M. Turek and K. A. Matveev, \textit{Cotunneling thermopower of single electron transistors}, Phys. Rev. B \textbf{65}, 115332 (2002).

\bibitem{MAprl} A. V. Andreev and K. A. Matveev, \textit{Coulomb blockade oscillations in the thermopower of open quantum dots}, Phys. Rev. Lett. \textbf{86}, 280 (2001).

\bibitem{MAprb} K. A. Matveev and  A. V. Andreev, \textit{Thermopower of a single-electron transistor in the regime of strong inelastic cotunneling}, Phys. Rev. B \textbf{66}, 045301 (2002).

\bibitem{Krawiec} M. Krawiec and K. I. Wysoki\'nski, \textit{Thermoelectric effects in strongly interacting quantum dot coupled to ferromagnetic leads}, Phys. Rev. B \textbf{73}, 075307 (2006).

\bibitem{Costi} T. A. Costi and V. Zlati\'c, \textit{Thermoelectric transport through strongly correlated quantum dots}, Phys. Rev. B \textbf{81}, 235127 (2010).

\bibitem{Trocha} P. Trocha and J. Barna\'s, \textit{Large enhancement of thermoelectric effects in a double quantum dot system due to interference and Coulomb correlation phenomena}, Phys. Rev. B \textbf{85}, 085408 (2012).

\bibitem{Donsa} S. Donsa, S. Andergassen, and K. Held, \textit{Double quantum dot as a minimal thermoelectric generator}, Phys. Rev. B \textbf{89}, 125103 (2014).

\bibitem{Wojcik} K. P. W\'ojcik and I. Weymann, \textit{Thermopower of strongly correlated T-shaped double quantum dots}, Phys. Rev. B textbf{93}, 085428 (2016).

\bibitem{Staring} A. A. M. Staring, L. W. Molenkamp, B. W. Alphenaar, H. van Houten, O. J. A. Buyk, M. A. A. Mabesoone, C. W. J. Beenakker and C. T. Foxon, \textit{Coulomb-blockade oscillations in the thermopower of a quantum dot}, Europhys. Lett. \textbf{22}, 57 (1993).

\bibitem{TE_exp1} A. S. Dzurak, C. G. Smith, C. H. W. Barnes, M. Pepper, L. Martin-Moreno, C. T. Liang, D. A. Ritchie and G. A. C. Jones, \textit{Thermoelectric signature of the excitation spectrum of a quantum dot}, Phys. Rev. B \textbf{55},  R10197(R) (1997).

\bibitem{TE_exp2} R. Scheibner, H. Buhmann, D. Reuter, M. N. Kiselev and L. W. Molenkamp, \textit{Thermopower of a Kondo spincorrelated quantum dot}, Phys. Rev. Lett. \textbf{95}, 176602 (2005).

\bibitem{TE_exp3} R. M. Potok, I. G. Rau, H. Shtrikman, Y. Oreg and D. Goldhaber-Gordon, \textit{Observation of the two-channel Kondo effect}, Nature \textbf{446}, 167 (2007).

\bibitem{Svilans} A. Svilans, M. Josefsson, A. M. Burke, S. Fahlvik, C. Thelander, H. Linke, and M. Leijnse, \textit{Thermoelectric characterization of the Kondo resonance in nanowire quantum dots}, Phys. Rev. Lett. \textbf{121}, 206801 (2018).

\bibitem{Dutta} B. Dutta, D. Majidi, A. G. Corral, P. A. Erdman, S. Florens, T. A. Costi, H. Courtois, and C. B. Winkelmann, \textit{Direct probe of the seebeck coefficient in a Kondo-correlated single-quantum-dot transistor}, Nano Lett. \textbf{19}, 506 (2019).

\bibitem{karki2020} D. B. Karki, \textit{Coulomb blockade oscillations of heat conductance in the charge Kondo regime}, Phys. Rev. B\textbf{102}, 245430 (2020).

\bibitem{CM} M. Cutler and N. F. Mott, \textit{Observation of Anderson Localization in an Electron Gas}, Phys. Rev. \textbf{181}, 1336 (1969).

\bibitem{thanharx} T. K. T. Nguyen and M. N. Kiselev, \textit{Generalized Cutler-Mott relation in a two-site charge Kondo simulator}, arXiv:2508.07891 (2025).

\bibitem{WF} R. Franz and G. Wiedemann, \textit{Ueber die W\"arme-Leitungsf\"ahigkeit der Metalle}, Anna. der Phys. \textbf{165}, 497 (1853). 

\bibitem{kiselev2023} M. N. Kiselev, \textit{Generalized Wiedemann-Franz law in a two-site charge Kondo circuit: Lorenz ratio as a manifestation of the orthogonality catastrophe}, Phys. Rev. B \textbf{108}, L081108 (2023).

\bibitem{shotnoise1} M. J. M. de Jong and C. W. J. Beenakker, Shot Noise in MesoscopicSystems, in \textit{Mesoscopic Electron Transport}, ed. L. L. Sohn, L. P.Kouwenhoven, and G. Sch\"on (Springer, Dordrecht, 1997) p. 225.

\bibitem{shotnoise2} Y. M. Blanter and M. B\"uttiker, \textit{Shot noise in mesoscopic conductors}, Phys. Rep. \textbf{336}, 1 (2000).

\bibitem{shotnoise3} T. Martin, Noise in Mesoscopic Physics, in \textit{Nanophysics: Coherence and Transport}, ed. H. Bouchiat, Y. Gefen, S. Gu\'eron, G.Montambaux, and J. Dalibard (Elsevier, Amsterdam, 2005).

\bibitem{LandauSela} L. A. Landau, E. Cornfeld, and E. Sela, \textit{Charge fractionalization in the two-channel Kondo effect}, Phys. Rev. Lett. \textbf{120}, 186801 (2018).

\bibitem{reviewGiazotto} F. Giazotto, T. T. Heikkil\"a, A. Luukanen, A. M. Savin, and J. P. Pekola, \textit{Opportunities for mesoscopics in thermometry and refrigeration: Physics and applications}, Rev. Mod. Phys. \textbf{78}, 217 (2006).

\bibitem{Meir2002} Y. Meir and A. Golub, \textit{Shot Noise through a Quantum Dot in the Kondo Regime}, Phys. Rev. Lett. \textbf{88}, 116802 (2002).

\bibitem{Golub} A. Golub, \textit{Shot noise near the unitary limit of a Kondo quantum dot}, Phys. Rev. B \textbf{73}, 233310 (2006).

\bibitem{Gogolin} A.O. Gogolin and A. Komnik, \textit{Full counting statistics for the Kondo dot in the unitary limit}, Phys. Rev. Lett. \textbf{97}, 016602 (2006); \textit{Towards full counting statistics for the Anderson impurity model}, Phys. Rev. B \textbf{73}, 195301 (2006).

\bibitem{shotnoisereview} K. Kobayashi and M. Hashisaka, \textit{Shot noise in mesoscopic systems: from single particles to quantum liquids}, J. Phys. Soc. Jpn. \textbf{90}, 102001 (2021).

\bibitem{Tnoise_exp1} E. S. Tikhonov, D. V. Shovkun, D. Ercolani, F. Rossella, M. Rocci, L. Sorba, S. Roddaro, and V. S. Khrapai, \textit{Local noise in a diffusive conductor}, Sci. Rep. \textbf{6}, 30621 (2016).

\bibitem{Tnoise_exp2} O. S. Lumbroso, L. Simine, A. Nitzan, D. Segal, and O. Tal, \textit{Electronic noise due to temperature differences in atomic-scale junctions}, Nature (London) \textbf{562}, 240 (2018).

\bibitem{Tnoise_exp3} E. Sivre, H. Duprez, A. Anthore, A. Aassime, F. D. Parmentier, A. Cavanna, A. Ouerghi, U. Gennser, and F. Pierre, \textit{Electronic heat flow and thermal shot noise in quantum circuits}, Nat. Commun. \textbf{10}, 5638 (2019).

\bibitem{Tnoise_exp4} S. Larocque, E. Pinsolle, C. Lupien, and B. Reulet, \textit{Shot Noise of a Temperature-Biased Tunnel Junction}, Phys. Rev. Lett. \textbf{125}, 106801 (2020).

\bibitem{Tnoise_exp5} R. A. Melcer, B. Dutta, C. Spansl\"att, J. Park, A. D. Mirlin, and V. Umansky, \textit{Absent thermal equilibration on fractional quantum hall edges over macroscopic scale}, Nat. Comm., \textbf{13}, 376 (2022).

\bibitem{Tnoise_exp6} A. Rosenblatt, S. Konyzheva, F. Lafont, N. Schiller, J. Park, K Snizhko, M. Heiblum, Y. Oreg, and V. Umansky, \textit{Energy Relaxation in Edge Modes in the Quantum Hall Effect}, Phys. Rev. Lett. \textbf{125}, 256803 (2020).

\bibitem{Tnoise_theo1} E. Zhitlukhina, M. Belogolovskii, and P. Seidel, \textit{Electronic noise generated by a temperature gradient across a hybrid normal metal-superconductor nanojunction}, Appl. Nanosci. \textbf{10}, 5121 (2020).

\bibitem{Tnoise_theo2} J. Rech, T. Jonckheere, B. Gr\'emaud, and T. Martin, \textit{Negative Delta-T Noise in the Fractional Quantum Hall Effect}, Phys. Rev. Lett. \textbf{125}, 086801 (2020).

\bibitem{Tnoise_theo3} M. Hasegawa and K. Saito, \textit{Delta-T noise in the Kondo regime}, Phys. Rev. B \textbf{103}, 045409 (2021).

\bibitem{Tnoise_theo4} H. Duprez, F. Pierre, E. Sivre, A. Aassime, F. D. Parmentier, A. Cavanna, A. Ouerghi, U. Gennser, I. Safi, C. Mora, and A.
Anthore, \textit{Dynamical Coulomb blockade under a temperature bias}, Phys. Rev. Res. \textbf{3}, 023122 (2021).

\bibitem{Tnoise_theo5} G. Zhang, I. V. Gornyi, and C. Sp\"ansl\"att, \textit{Delta-T noise for weak tunneling in one-dimensional systems: Interactions versus quantum statistics}, Phys. Rev. B \textbf{105}, 195423 (2022).

\bibitem{detect_exp1} R. De-Picciotto, M. Reznikov, M. Heiblum, V. Umansky, G. Bunin, and D. Mahalu, \textit{Direct observation of a fractional charge}, Nature (London) \textbf{389}, 162 (1997).

\bibitem{detect_exp2} L. Saminadayar, D.C. Glattli, Y. Jin, and B. Etienne, \textit{Observation of the $e/3$ fractionally charged Laughlin quasiparticle}, Phys. Rev. Lett. \textbf{79}, 2526 (1997).

\bibitem{LeHur} C. Mora, X. Leyronas, and N. Regnault, \textit{Current noise through a Kondo quantum dot in a SU$(N)$ Fermi liquid state}, Phys. Rev. Lett. \textbf{100}, 036604 (2008); P. Vitushinsky, A. A. Clerk, and K. Le Hur, \textit{Effects of Fermi liquid interactions on the shot noise of an SU$(N)$ Kondo quantum dot}, Phys. Rev. Lett. \textbf{100}, 036603 (2008).

\bibitem{comment} In standard literature, the terms shot noise and delta-T noise typically refer to charge current fluctuations under a voltage bias and a temperature gradient, respectively. In this work, we also investigate heat current fluctuations and the cross-correlations between charge and heat currents under either a voltage bias or a temperature difference. Accordingly, we refer to these as voltage-driven electric noise, voltage-driven heat noise, voltage-driven mixed noise, and temperature-driven electric noise, temperature-driven heat noise, and temperature-driven mixed noise.

\bibitem{thanhcom2023} A. V. Parafilo and T. K. T. Nguyen, \textit{Thermopower of a Luttinger-liquid-based two-channel charge Kondo circuit: nonperturbative solution}, Commun. Phys. \textbf{33}, 1 (2023).

\bibitem{thanh2024} T. K. T. Nguyen, H. Q. Nguyen, and M. N. Kiselev, \textit{Thermoelectric transport across a tunnel contact between two charge Kondo circuits: Beyond perturbation theory},  Phys. Rev. B  \textbf{109}, 115139 (2024). 

\bibitem{Giamarchi} T. Giamarchi, \textit{Quantum Physics in One Dimension} (Oxford University Press, New York, 2004).

\bibitem{Aleiner} I. L. Aleiner and L. I. Glazman, \textit{Mesoscopic charge quantization}, Phys. Rev. B \textbf{57}, 9608 (1998).

\bibitem{Onsager} L. Onsager, \textit{Reciprocal Relations in Irreversible Processes. I.}, Phys. Rev. \textbf{37}, 405 (1931); \textit{Reciprocal Relations in Irreversible Processes. II.}, Phys. Rev. \textbf{38}, 2265 (1931).

\bibitem{Jauhobook} H. Haug , A. -P. Jauho, \textit{Quantum Kinetics in Transport and Optics of Semiconductors} (Solid-State Sciences, Vol. 123, Springer, 2008).

\bibitem{Mahanbook} G. D. Mahan, \textit{Many-Particle Physics} (Springer Science and Business Media, 2000).

\bibitem{thanh2015} T. K. T. Nguyen and M. N. Kiselev, \textit{Protection of a non-Fermi liquid by spin-orbit interaction}, Phys. Rev. B \textbf{92}, 045125 (2015). 

\bibitem{thanhcom2024} T. K. T. Nguyen, T. B. Cao, T. A. Chu, T. L. H. Nguyen, H. Q. Nguyen, and M. N. Kiselev, \textit{Effects of asymmetry in Kondo channels on thermoelectric efficiency}, Commun. Phys. \textbf{34}, 317 (2024).

\bibitem{kiselev2025} M. N. Kiselev, \textit{Universal scaling functions for a quantum transport through single-site and double-site charge Kondo circuits}, Lecture notes in Physics, in preparation.

\bibitem{PaKi2025a}  Andrei I. Pavlov and Mikhail N. Kiselev,\textit{
Universal relations between thermoelectrics and noise in mesoscopic transport across a tunnel junction},
ArXiv: 2508.05413 (2025).

\bibitem{PaKi2025b}  Andrei I. Pavlov and Mikhail N. Kiselev,\textit{
Noise signatures of a charged Sachdev-Ye-Kitaev dot in mesoscopic transport},
ArXiv: 2508.13098 (2025).

\end{thebibliography}
\end{document}